\documentclass[fleqn,usenatbib]{mnras}

\usepackage{newtxtext,newtxmath}
\usepackage{xcolor}

\usepackage[T1]{fontenc}
\usepackage{ae,aecompl}

\usepackage{graphicx}	
\usepackage{amsmath}	
\usepackage{amssymb}	

\graphicspath{{./}{figures/}}

\title[Star Formation in AGN]{Star Formation in Accretion Disks and SMBH Growth}

\author[A. J. Dittmann and M. C. Miller]{
Alexander J. Dittmann,$^{1}$\thanks{E-mail: \href{mailto:dittman@astro.umd.edu}{dittmann@astro.umd.edu}}
M. Coleman Miller,$^{1}$
\\
$^{1}${Department of Astronomy and Joint Space-Science Institute, University of Maryland, College Park, MD 20742-2421}}

\date{Accepted XXX. Received YYY; in original form ZZZ}

\pubyear{20XX}

\begin{document}
\label{firstpage}
\pagerange{\pageref{firstpage}--\pageref{lastpage}}
\maketitle

\begin{abstract}
Accretion disks around active galactic nuclei are potentially unstable to star formation at large radii.  We note that when the compact objects formed from some of these stars spiral into the central supermassive black hole, there is no radiative feedback and therefore the accretion rate is not limited by radiation forces. Using a set of accretion disk models, we calculate the accretion rate onto the central supermassive black hole in both gas and compact objects. We find that the timescale for a supermassive black hole to double in mass can decrease by factors ranging from $\sim0.7$ to as low as $\sim0.1$ in extreme cases, compared to gas accretion alone. Our results suggest that the formation of extremely massive black holes at high redshift may occur without prolonged super-Eddington gas accretion or very massive seed black holes. We comment on potential observational signatures as well as implications for other observations of active galactic nuclei.
\end{abstract}




\section{Introduction}
The existence of supermassive black holes (SMBH) with masses $\gtrsim10^9~M_\odot$ at redshifts $z \gtrsim 7$ \citep[e.g.,][]{2011Natur.474..616M}
challenges our understanding of the formation of SMBH seeds and their subsequent growth.  The fundamental
issue is that even if stellar-origin black holes with masses $M\sim
10-100~M_\odot$ form from the first stars at $z\sim 20-30$, there is not
enough time to reach $M\sim 10^9~M_\odot$ at $z\sim 7$ if accretion is
Eddington-limited at a standard black hole accretion efficiency
$\eta\equiv L/({\dot M}c^2)\approx 0.1$ for a luminosity $L$ and an
accretion rate ${\dot M}$. Quantitatively, the exponential growth time
for Eddington-luminosity accretion at efficiency $\eta$ is $\tau\approx
4.5\times 10^7~{\rm yr}\left(\eta/0.1\right)$.  Thus in the $\sim 500$~Myr
between $z=20$ and $z=7$ the growth factor is only
$\sim 6\times 10^4$ for $\eta=0.1$.  This problem is usually solved by
invoking massive seed black holes from direct collapses \citep{2006MNRAS.370..289B}, Population
III stars \citep{2013RPPh...76k2901B}, or mergers of stars or black holes in dense clusters
\citep{{2009ApJ...694..302D},{2011ApJ...740L..42D}}; or by supposing that gas accretion can proceed at a few times the
standard Eddington rate \citep{2019MNRAS.483.2031T}.

Here we consider a different scenario, in which stellar-origin black holes
in gas surrounding a massive black hole migrate inward and merge with the
central black hole.  These mergers produce very little radiation, and
therefore the effective $\eta$ is decreased.  A
decrease in $\eta$ of only an average factor of $\sim 0.5$ would mean that there is
enough time for a hole to grow from stellar masses to $\sim 10^9~M_\odot$
by $z\sim 7$.

In more detail, analytical models of accretion disks often predict disks unstable to
gravitational perturbations at large radii. Disks become unstable when the Toomre criterion \citep{1964ApJ...139.1217T}
is satisfied, 
$Q\equiv c_s\kappa_e/(\pi G\Sigma) \lesssim 1$ for a gas disk, where $c_s$ is the sound speed,
 $\kappa_e$ is the radial epicyclic frequency, $\Sigma$ is the surface density, and $G$ is Newton's constant.
In a standard accretion disk supported by gas pressure with opacity dominated by free-free absorption \citep{1973A&A....24..337S},
$Q\propto r^{-9/8}$, so gravitational instability is expected at large radii. If the cooling timescale in the disk is shorter than the dynamical timescale for the disk, gravitational instability can 
lead to disk fragmentation into dense objects such as stars or planets \citep{2001ApJ...553..174G}.
Gravitational fragmentation has also been observed in both smoothed particle and Eulerian hydrodynamics simulations
(e.g., \citealt{{2007MNRAS.379...21N},{2011ApJ...730...45J}}).  

Considering fragment masses and balancing stellar accretion and mass loss, typical stellar masses in the disk range from 50-500 $M_\odot$, which is unlikely to be massive enough to open a gap in the accretion disk.
If individual stars and black holes are not massive enough to open gaps in the disk, they migrate due to Lindblad and corotation torques with
the disk, resulting in inwards migration that can be much faster than the viscous timescale of the gas
\citep{{2002ApJ...565.1257T},{2008A&A...485..877P}}. 
Once the objects are  close enough to the central SMBH, 
torques from gravitational radiation \citep{1964PhRv..136.1224P} dominate, leading to rapid mergers regardless of gap opening.
These mergers can grow an SMBH without limitation by radiation forces, facilitating faster growth 
than gas accretion alone. 

Star formation in AGN disks has been investigated before \citep[e.g.,][]{{1980SvAL....6..357K},{2003ApJ...590L..33L},{2004ApJ...608..108G},{2007MNRAS.374..515L}}, although our work differs in our choice of disk model and use of updated opacities in the star-forming region of the disk. Work by \citet{2016ApJ...828..110I} has explored the implications of star formation in AGN disks on limiting maximum SMBH masses. The migration of small numbers of stellar-mass black holes has been examined by \citet{2019ApJ...878...85S}, and the implications of stellar-mass black hole mergers in AGN disks have been investigated from a variety of angles
\citep[e.g.,][]{{2012MNRAS.425..460M},{2014MNRAS.441..900M},{2017ApJ...835..165B},{2017MNRAS.464..946S},{2018ApJ...866...66M},{2019MNRAS.490L..42F},{2019arXiv190704356M}}.
We in turn study the implications of star formation on the growth of high-z SMBHs. In particular, the understanding of migration has changed significantly over the last decade (see Section \ref{subsec: migration}), whereas many previous studies have relied on more simplistic migration prescriptions that can produce migration rates differing by orders of magnitude.

In this work we 
investigate a different regime of star formation in AGN disks.
We use a series of star-forming steady state disk models closely following those described in \citet{2005ApJ...630..167T}(hereafter TQM). 
These models connect a star-forming outer disk to a gravitationally stable inner disk. 
The remainder of the paper is divided as follows: In Section \ref{sec:methods} we summarise results that we use in our analysis, 
comment on various assumptions, and provide a detailed description of our disk models. In Section \ref{sec:Results} and Section \ref{sec:Discussion} we present the results of our analyses and discuss implications for observations. Our conclusions are in Section \ref{sec: conclusion}. 

\section{Methodology} \label{sec:methods}
In this section we present a number of results that we use throughout our analysis, as well as the details of our disk models. 
Because of the uncertainties in many relevant aspects of astrophysics, we must make many approximations throughout this analysis. We therefore review our choices, and emphasise
that although they influence the fine details of our results, our qualitative 
conclusions do not depend critically on our assumptions. 

\subsection{Disk Instability} \label{subsec:instability}
When Q $\lesssim1$, disks are thought to evolve towards stability by processes such as gravitational collapse or by transport of angular momentum via global torques such as bars or spiral waves. 
The immediate evolution of a gravitationally-unstable disk is determined by the cooling and dynamical timescales. 
If the cooling timescale ($t_c$) is sufficiently short compared to the dynamical timescale ($t_{\rm dyn} \equiv 1/\Omega$, where $\Omega$ is the orbital angular velocity), $t_c \lesssim 3t_{\rm dyn}$, clumps of gas
may continue to collapse into dense objects \citep{2001ApJ...553..174G} such as stars or planets. Cooling timescales predicted by AGN disk models are usually sufficiently 
short to facilitate the collapse of gas into dense objects. AGN disk models contrast with those describing circumstellar disks 
\citep[e.g.,][]{2006ApJ...636L.149C}, where core-accretion is thought to be the dominant form of initial compact 
object growth \citep{1986Icar...67..391B}. We note that $t_c \gtrsim 50 ~t_{\rm dyn} $
may be necessary to suppress disk fragmentation in the presence of
turbulence, such as that caused by the magneto-rotational instability \citep{2013ApJ...776...48H}, so our adopted conditions for fragmentation
may be overly stringent.

Although the conditions $Q \lesssim 1$ and $t_c \lesssim 3t_{\rm dyn}$ may be sufficient to determine 
whether disks fragment, it is not clear whether the resulting fragments are massive enough to become stars. 
The initial mass of fragments in the disk may be approximated roughly by the Jeans mass, $M_J \sim c_s^3 \rho^{-1/2}$, where $\rho$ is the disk gas density. If cooling is efficient then as the initial fragments collapse $c_s^3 \rho^{-1/2}$ gradually decreases. As long as the thermal adjustment timescale is shorter than the free fall timescale for the fragment, the collapse is approximately isothermal. Because density increases during collapse and $c_s^2=k_bT/\mu$, where T is gas temperature, $k_b$ is the Boltzmann constant, and $\mu$ is the mean particle mass, one expects each clump to fragment repeatedly until the assumption of isothermal collapse breaks down. 
The minimum fragment mass can be changed by other physics, such as the presence of magnetic fields.

To make a rough estimate of the final fragment masses in our disk models, we apply the result of \citet{1976MNRAS.176..367L}. The minimum Jeans mass is given by
\begin{equation}
M_{\rm frag}\approx1.54\times10^{-3}T^2_b\frac{\kappa_f}{\kappa_0}~ M_\odot, 
\label{eqn: frag_mass}
\end{equation}
where $\kappa_0$ is the electron scattering opacity, $\kappa_f$ is the final opacity of the fragment, and $T_b$ is the effective temperature of the disk in Kelvin. This is appropriate for our disk models because fragments will be bathed in thermal radiation from the disk, and the disks in our models are optically thick to their own radiation. In the regions of our disks where we expect star formation, temperatures are usually between 100 and 1000 K and opacities only vary by factors of order unity. Accordingly, we assume that $\kappa_f$ is the same as the initial opacity before collapse. Our predicted stellar masses are significantly larger than those for molecular clouds, largely because the accretion disks we consider are significantly hotter than present day molecular clouds.

Both our predicted $M_{\rm frag}$ (e.g., $\sim 150~ M_\odot$ for $T_b\sim 200 ~\rm K, \kappa \sim 1~cm^2/g$) and our estimates of stellar mass based on accretion and mass loss arguments (see Section \ref{subsec:accwind}) produce mass estimates of order $\sim 200~ M_\odot$. An upper limit on the timescale for the collapse of protostars of mass $m > 100 ~M_\odot$ into stars can be obtained from the Kelvin-Helmholtz timescale \citep{{1984ApJ...280..825B},{2004ApJ...608..108G}}, 
\begin{equation}\label{eqn:tkh}
t_{\rm KH}\approx 3300\left(\frac{\kappa}{0.4\rm ~cm^{2} ~g^{-1}}\right)~\rm yr,
\end{equation}
where $\kappa$ is the opacity of the gas in the disk, we have dropped the weak dependence on mass, and we use solar metallicities. This timescale is shorter than one might think, based on less massive stars, because massive stars have smaller specific binding energies.  
Since we expect star formation to occur at distances from the SMBH on the order of parsecs, where $t_{\rm dyn}\sim 10^4~{\rm yr}$ for a $4\times10^6~M_\odot $ SMBH,  $t_{\rm KH}$ is a few times larger than $t_{\rm dyn}$. 
We use 
\begin{equation}
\Omega^2 = \frac{G M_\bullet}{r^3} + \frac{2\sigma^2}{r^2}, 
\end{equation}
where $M_\bullet$ is the SMBH mass and $\sigma$ is the stellar velocity dispersion for stars near the SMBH but not in the disk. $\Omega^2 \sim G M_\bullet/r^3$ is reasonably accurate for our approximate calculations. We show in Section \ref{sec:Results} that $t_{\rm KH}$ is orders of magnitude less than the characteristic timescale for stellar growth via accretion in the outer regions of the disk, which indicates that stars do not accumulate significant additional mass before beginning fusion on the main sequence. 

\subsection{Migration} \label{subsec: migration}
Stellar-mass black holes in AGN disks migrate analogously to planets in circumstellar disks. Our disk models have opening angles $H/r \sim 0.01 - 0.2$. 
We assume that stars and black holes form in almost circular orbits with inclinations $i\sim H/r$ and find that stellar masses are less than $\sim 10^3~M_\odot$. Then, $\langle e^2 \rangle+\langle i^2\rangle \gtrsim (2m/M_\bullet)^{2/3}$, where $m$
is the mass of the star, $e$ is the orbit eccentricity, and $M_\bullet$ is the mass of the central SMBH, $4\times10^6~M_\odot$, so the collisional evolution of compact objects in the disk is
dispersion-dominated \citep{2010AJ....139..565R}. In this regime, eccentricities and inclinations follow a Rayleigh distribution \citep{1992Icar...96..107I},
\begin{equation}
f(e^2,i^2) \propto \frac{1}{\langle e^2 \rangle \langle i^2\rangle}\exp\left(-\frac{e^2}{\langle e^2\rangle}-\frac{i^2}{\langle i^2\rangle}\right),
\end{equation}
which exponentially suppresses highly inclined or eccentric orbits as the system evolves. The assumption that $i\sim H/r$ may be an overestimate if star formation occur preferentially towards the midplane. In either case, it is reasonable even in the absence of damping to assume that orbits are initially nearly circular and coplanar. 

Collisions between compact objects excite eccentricities and inclinations over time \citep{2000Icar..143...28S}.
However, orbits passing through the gas disk excite waves which damp orbital eccentricities and inclinations \citep{2004ApJ...602..388T}.
\citet{2008A&A...482..677C} investigated the competition between these two effects using hydrodynamic and N-body simulations of eight protoplanets, and 
concluded that damping forces dominated collisions, leading to nearly coplanar and circular orbits. However, this simulation included a relatively small number of objects, and the conclusion may not hold for systems including more objects, in which the frequency of interactions increases dramatically. For the current work we assume that this result also holds for much larger numbers of disk-embedded objects. 

It is not clear if increases in inclination and eccentricity over time due to close gravitational encounters would significantly change migration timescales, especially if the overall growth timescale for $i$ or $e$ is large compared to migration timescales. However, \citet{2000MNRAS.315..823P} and  \citet{2008A&A...482..677C} find that nonzero eccentricity and inclination can decrease inward migration rates, even reversing the sign of the torque for highly eccentric orbits. Larger eccentricities also lead to greater gravitational radiation torques, so it is not clear how overall migration timescales would be affected. Eccentricity and inclination damping become weaker near the SMBH because the damping timescale depends more strongly on radius than the dynamical timescale, $t_{\rm damp}/t_{\rm dyn}\propto r^{-2}$ for constant $H/r$ \citep{2004ApJ...602..388T}, which may lead to significant eccentricities in the regime where gravitational radiation torques dominate. 

We consider multiple torques that lead to migration of compact objects through the disk. Regardless of disk structure, compact objects orbiting the SMBH will lose angular momentum to gravitational radiation at an average rate \citep{1964PhRv..136.1224P}
\begin{equation}
\Gamma_{\rm GW}=-\frac{32}{5}\frac{G^{7/2}m^2M_\bullet^2(m+M_\bullet)^{1/2}}{c^5a^{7/2}(1-e^2)^2}\left(1+\frac{7}{8}e^2\right).
\end{equation}

Torques on the migrating object from the disk depend on whether the object is massive enough to open a gap. The criterion for gap opening is 
approximately given by $g\lesssim1$, where
\begin{equation}
g=\frac{3}{4}\frac{H}{r}\left(\frac{q}{3}\right)^{-1/3}+50\frac{\alpha}{q}\left(\frac{H}{r}\right)^2,
\label{eq:g}
\end{equation}
\citep{2011ApJ...726...28B} where $\alpha$ is the usual viscosity parameter \citep{1973A&A....24..337S} and $q=\frac{m}{M_\bullet}$. 
Here, the first term represents the balance between the size of an object's Hill sphere and the disk scale height, and the second term represents how the gap may be prevented from opening by viscosity. 

We consider different migration torques 
depending on whether $g$ is greater than or less than 1, although in our disk models gap opening is relatively rare, often happening only for objects with $q \gtrsim 10^{-4} - 10^{-3}$ depending on the disk model (Figure \ref{fig: migration_time}). Throughout most of the disk the first term in Equation (\ref{eq:g}) is less than unity, so in practice the viscosity often plays the deciding role. We denote the minimum mass at a given radius which is able to open a gap as the isolation mass $M_{\rm iso}$.

When no gap is present, we calculate a migration torque using the expressions for Lindblad and non-linear corotation torques found by \citet{2010MNRAS.401.1950P}.
For azimuthally isothermal disks the normalised torque is 
\begin{equation}
\begin{split}
\frac{\Gamma_{\rm iso}}{\Gamma_0}=1.1\psi-0.9\beta-\delta-2.5 \\
\psi\equiv-\frac{d \ln\Omega}{d\ln r}, \beta\equiv-\frac{d\ln T}{d\ln r}, \delta\equiv-\frac{d \ln\Sigma}{d\ln r},
\end{split}
\label{eqn: t_iso}
\end{equation}
where $\Sigma$ is the disk surface density and $\Gamma_0=(q/H)^2\Sigma r^4\Omega^2$.
Note that this notation is slightly different than used by \citet{2010MNRAS.401.1950P}, because we use $\alpha$ to refer to disk viscosity and we assume a rotation profile with deviations from Keplerian motion. 
For azimuthally adiabatic disks with adiabatic index $\gamma$, the normalised torque is given by
\begin{equation}
\frac{\gamma\Gamma_{\rm ad}}{\Gamma_0}=1.1\psi-1.7\beta-\delta+7.9\frac{\beta-(\gamma-1)\delta}{\gamma}-2.5.
\label{eqn: t_adi}
\end{equation}
When applying these formulae, we use $\gamma=5/3$, and interpolate between the two expressions following \citet{2010ApJ...715L..68L} to find the
total gas torque:
\begin{equation}
\Gamma_g=\frac{\Gamma_{\rm ad}\Theta^2+\Gamma_{\rm iso}}{(\Theta+1)^2},
\end{equation}
where 
\begin{equation}
\Theta = \frac{c_v\Sigma\Omega\tau_e}{12\pi\sigma_{\rm sb}T^3}
\end{equation}
relates the dynamical and cooling timescales, $c_v = 1.5k_b$, $\sigma_{\rm sb}$ is the Stefan-Boltzmann constant, and $\tau_e$ is the effective optical depth at the disk midplane. From \citet{1990ApJ...351..632H}
\begin{equation}
\tau_e = \frac{3\tau}{8}+\frac{\sqrt{3}}{4}+\frac{1}{4\tau}.
\end{equation}
The optical depth is $\tau=\kappa\rho H$.
In principle, the total gas torque could be positive in some regions of the disk, which would push objects outwards and thus prevent inwards migration \citep{2010ApJ...715L..68L}. \citet{2016ApJ...819L..17B} identify these `migration traps' in some models of AGN disks. For each of our models, we have verified that the total torque, $\Gamma_{\rm tot}=\Gamma_g+\Gamma_{\rm GW}$ on migrating objects is never positive, and thus migration only proceeds inwards. We plot the total torque on a migrating object with mass $m=10~M_\odot$ as a function of radius in Figure \ref{fig: disk_par}. 

If a gap opens in the disk, migration changes significantly. Historically, gas was believed to be unable to cross gaps in the disk, and migration was thought 
to proceed at the rate of viscous gas inflow. However, numerous recent numerical studies have demonstrated that gas is able to flow through gaps on horseshoe orbits \citep{{2014ApJ...792L..10D},{2015A&A...574A..52D}}, and that migration is tied to the density of gas in the gap \citep{2018ApJ...861..140K}. In situations where gap opening occurs, we use the approximate radial migration formula of \citet{2018ApJ...861..140K} in terms of the viscous inflow velocity $v_{\rm vis}=-\psi \alpha H^2\Omega/R$
\begin{equation}
v = 100v_{\rm vis}\Sigma r^2/m.
\label{eqn: gap_migration}
\end{equation}
We find that when gaps open at radii where gravitational radiation torques are insignificant, migration timescales increase by orders of magnitude, as shown in Figure \ref{fig: migration_time}.

We do not include runaway type-III migration, in which objects are massive enough to open a partial gap which is front-back asymmetric \citep{{2008MNRAS.386..179P},{2003ApJ...588..494M}}. Such migration could temporarily increase migration rates, but would likely be rare and short-lived.
It is also possible that multiple massive migrating object could open gaps in the disk and fall into resonances. Investigation into the effects of mean motion resonances on the orbits of migrating objects in AGN disks is beyond the scope of this work, but if gap-opening migrators in mean motion resonances move outwards, then lower mass objects could be scattered inwards, much closer to the SMBH, similarly to models of the Solar System \citep{2011Natur.475..206W}.
We also neglect diffusive migration due to turbulence
\citep{2006ApJ...647.1413J} since we expect most stars to have sufficient mass that diffusive migration plays a negligible role. Additionally, we have neglected the effects of accretion and feedback on migration torques. Two-dimensional disk simulations have suggested that feedback or efficient accretion could alter torques \citep{2019MNRAS.486.2754D}, but we neglect these effects at present because they are insufficiently understood. For similar reasons, we neglect the interactions of spiral density waves from multiple objects.  

\subsection{Stellar Accretion and Winds} \label{subsec:accwind}

We assume that stars form in disks as described in Section \ref{subsec:instability}. The outer regions of our disk models have fairly large opening angles, $H/r \sim 0.2$, so the mass ratio required to open a gap is quite large. We verify this in Section \ref{sec:Results}. Since stellar masses are, at least initially, too small to affect disk structure in major ways, we approximate accretion onto stars by the Bondi accretion rate 
\begin{equation}\label{eqn:bondi}
\dot{m}_B \approx \pi \rho G^2m^2/c_s^3.
\end{equation}
We approximate the time required for an object to double in mass via Bondi accretion as $t_2 \approx c_s^3/(2\pi G^2 m\rho) $. 
Because in a quasi-stable disk with $Q\sim 1$, gas density decreases sharply with increasing distance from the SMBH, this accretion rate is quite small in the outer regions of the accretion disk. However, at radii within a parsec from the SMBH, the doubling timescale becomes comparable to the dynamical timescale.

Since the Bondi accretion rate is proportional to $m^2$, this accretion rate can become unphysically large as mass increases. In such cases, we assume that the accretion rate onto stars is limited such that the accretion luminosity, $L_{\rm acc}=Gm\dot{m}_{\rm edd}/r_*$ (for stellar radius $r_*$) is less than the Eddington luminosity, $L_{\rm edd} \approx 1.3 \times 10^{38}(m/M_\odot)~ \rm erg ~ s^{-1}$. 
Thus, when the opacity is dominated by electron scattering, the accretion rate is regulated to be less than 
\begin{equation}
\dot{m}_{\rm edd} \approx 10^{-3}(r_*/r_\odot) ~M_\odot ~ \rm yr^{-1}
\end{equation}
(see \citet{1993ApJ...409..592A} for a similar treatment). By following the procedure in \citet{1984ApJ...280..825B} to determine the radius of a massive star, using solar metallicity, we find that $r_*/r_\odot$ can be approximated well by $r_*/r_\odot = 0.56(m/M_\odot)^{0.51}$ in the 100-1000 $M_\odot$ range. We can thus approximate the Eddington-limited accretion rate onto stars as
\begin{equation}\label{eqn:edd}
\dot{m}_{\rm edd} = 5.6\times10^{-4}\left(\frac{0.4 ~\rm cm^2 ~\rm g^{-1}}{\kappa}\right)\left(\frac{m}{M_\odot}\right)^{0.51}~  M_\odot ~ \rm yr^{-1}.
\end{equation}

Bondi accretion itself can be modified by the stellar radiation. The force on infalling matter is reduced by a factor of $(1-f)$, where $f$ is the ratio of local flux to the Eddington flux. If the accretion flow is spherically symmetric and optically thin out to the Bondi radius, then the reduction factor is $(1-\Gamma)$, where $\Gamma$ is the ratio of the stellar luminosity to the Eddington luminosity. We approximate the results of \citet{1984ApJ...280..825B} according to a power law
\begin{equation}\label{eqn:eddfact}
(1-\Gamma)\approx 4.2(m/M_\odot)^{-0.43}.
\end{equation}
Because mass is conserved, ${\dot m}\propto r_B^2\rho u$, where $u$ is the speed of the inflow at large distances and $r_B\propto m$ is the radius inside of which gravity controls the flow.  For this purpose, the factor $(1-\Gamma)$ acts to reduce the effective mass of the object, and therefore the accretion rate becomes
\begin{equation}\label{eqn:bondi2}
\dot{m}_B \approx \pi \rho G^2m^2(1-\Gamma)^2/c_s^3.
\end{equation}

Because the optical depth $\tau_B=\kappa(Gm/c_s^2)\rho$ is less than unity in the outer regions of our disk models ($r>1$~pc) for stars with $m\sim 100~M_\odot$, and because both radiative and gravitational accelerations scale as $r^{-2}$, we expect $\Gamma$ to be nearly constant throughout the entire Bondi region in the outer disk.  However, as $\rho$ and $r_B$ increase at radii closer to the SMBH, $\tau_B$ can be greater than unity depending on the opacity of the disk.  At smaller radii, differential rotation of the disk at either side of a star's Bondi radius can also reduce the Bondi accretion rate.  Moreover, if accretion onto stars deviates significantly from spherical symmetry, the impact of this mechanism may be reduced.  However, we suggest that in these scenarios, stellar masses will be limited by enhanced mass loss, which we now describe.

We can estimate mass loss rates of O stars using the empirically-calibrated result by \citet{1993ApJ...412..771L}
\begin{equation}
\rm log(\dot{M})=1.738~\rm log (L)-1.352~ \rm log (T_{\rm eff})-9.547,
\end{equation}
where $\dot{M}$ is measured in $M_\odot ~ \rm yr^{-1}$, $L$ is measured in $L_\odot$, and $T_{\rm eff}$ is measured in Kelvin. 
We use the result from \citet{1984ApJ...280..825B} that very massive Population I and II stars have effective temperatures 
of $\sim 6\times10^4~K$, almost independent of mass. Then, we approximate $L/L_\odot \approx 5.5\times10^3(m/M_\odot)^{1.2}$ in the range $100-1000~ M_\odot$ based on the expression for $L/L_{\rm edd}$ in \citet{1984ApJ...280..825B}. This enables us to estimate the mass loss rate as 
\begin{equation}\label{eqn:winds}
\dot{M}_{\rm loss}\approx 3.1\times10^{-10}(m/M_\odot)^{2.1} ~M_\odot ~ \rm yr^{-1}.
\end{equation}

Additionally, stellar rotation can strongly enhance wind mass loss rates. We expect that all stars born inside the disk will have some rotational angular momentum, potentially enhancing these mass loss rates. The outer portions of our disk models have a large  opening angle $H/r\sim0.2$, so we do not expect mass loss to change by factors of more than order unity. However, the thinner inner star-forming regions of the disk have $H/r\sim0.01$. Considering the Hill radius of migrating objects, $r_H/r=(m/3M_\bullet)^{1/3}$, and a fiducial black hole mass of $4\times10^6 ~M_\odot$, objects in the disk with mass greater than $\sim 10 ~M_\odot$ would experience primarily aspherical accretion. Additionally, as distance from the SMBH decreases, the difference in velocity between the outward and inward (with respect to the SMBH) Bondi radius of the migrating black hole increases due to increases in both the disk angular velocity and the Bondi radius. 

It is necessary to understand the balance of accretion and mass loss to make good estimates of the mass of stars at the end of their lives. We can approximate the final masses of stars in the disk by balancing these processes. We calculate the equilibrium mass as the mass for which Equations (\ref{eqn:winds}) and either (\ref{eqn:edd}) or (\ref{eqn:bondi2}) balance. In practice, attempting to solve directly for mass using Equations (\ref{eqn:bondi2}) and (\ref{eqn:winds}) can result in a mass lower than was assumed in writing these equations, so in these cases we expect equilibrium masses to be between 50 and 100~ $M_\odot$.

However, our estimates for final masses are likely overestimates, since we have neglected enhanced mass loss rates towards the end of stellar lifetimes, and also neglected mass loss due to pulsational instability in stars with $m \gtrsim 100~M_\odot$. Mass loss driven by pulsations can range from $\sim 10^{-6}-10^{-4}~ M_\odot ~ \rm yr^{-1}$ for stars with masses $100 \lesssim m \lesssim 200 ~M_\odot$ 
\citep{{1970A&A.....5..355A},{1973MNRAS.162..169P}}. 
Additionally, the aforementioned mass loss rates do not depend explicitly on rotation. For example,
\citet{2000A&A...361..159M} found analytically that mass loss rates can become asymptotically large for stars with moderate rotation and $\Gamma \gtrsim 0.64$. Thus, comparing with Equation (\ref{eqn:eddfact}),
we expect maximum masses ranging from $300 - 500 ~M_\odot$, if our other considerations would predict larger masses. Balancing Equations (\ref{eqn:winds}) and (\ref{eqn:edd}) informs us that the equilibrium Eddington-limited stellar mass is
\begin{equation}
m_{\rm edd}/M_\odot \approx \left(1.8\times10^6 \left(\frac{0.4 ~\rm cm^2 ~\rm g^{-1}}{\kappa}\right) \right)^{1/1.59}\sim 10^4, 
\end{equation}
so we expect stars to become limited by violent mass loss before they approach the Eddington limit. 

We note that some regions of the disk, the final stellar masses are in the expected range for pair-instability supernovae. For example, \citet{2012ApJ...748...42C} find that pair
instability supernovae occur for stars as small as $65 ~M_\odot$, although core collapse also occurred in some of their models at each mass until $80 ~ M_\odot,$ after which either pair-instability supernovae or pulsational pair-instability supernovae are possible. However,
\citet{2001ApJ...550..372F} found that the pair-instability explosion was unable to unbind their
$300~M_\odot$ model, resulting in black hole formation. Thus, we expect some violent transients from the stars formed in our disk models, but expect that many black holes will be left behind.

\subsection{Disk Model} \label{subsec:models}

In our investigation, we consider the TQM accretion disk models along with the modifications to the opacity 
prescription described below. These models connect an outer accretion disk, where radiation pressure from massive stars or accretion onto black holes provides the support necessary to maintain $Q\sim1$, to an inner gravitationally stable $\alpha$-disk. We find this model advantageous because it provides a straightforward way to track the gas lost from the disk during star formation. Additionally, since the model includes the effects of irradiation on disk structure, this may lead to more realistic inferences of initial fragment masses, accretion rates, and migration rates. As long as irradiation does not increase by orders of magnitude once objects migrate into the gravitationally stable region, the pressure support of the inner disk is not significantly modified by the migration of objects through it.

The TQM disk models have a number of free parameters. For our purposes, the most important are the viscosity parameter $\alpha$, the SMBH mass, the accretion rate at the outer boundary of the disk ($\dot{M}_{\rm out}$), and the efficiency with which rest mass energy from star formation is converted into radiation ($\epsilon$). We consider accretion onto a $4\times10^6~ M_\odot$ SMBH. In principle, disks around lower mass black holes can also be gravitationally unstable, but we choose not to investigate these since the relation between central black hole mass and stellar bulge velocity dispersion is poorly measured for black holes less massive than $\sim10^6~ M_\odot$ \citep{2009ApJ...698..198G}. On the other side of the spectrum, accretion disks around more massive black holes are more susceptible to gravitational instability \citep[see, e.g.,][]{2016ApJ...828..110I}, although because we are concerned with SMBH growth in the early universe, it is pertinent to study lower mass SMBHs. 

We assume a supernova feedback parameter $\xi=1$, following TQM. 
Here, $\xi$ is a dimensionless parameter representing non-radiative pressure support due to feedback that is independent of optical depth. 
In principle, extreme mass loss from stars or stellar-mass black hole-driven outflows could increase this value significantly, causing $\xi=1$ to overestimate the star formation rate and underestimate the gas accretion rate onto the SMBH. Since the input physics is highly uncertain, we parametrise this by considering how our results change if we have incorrectly estimated accretion rates in this manner. We also assume a stellar velocity dispersion $\sigma=180(M_\bullet/2\times10^8 ~M_\odot)^{0.23}~{\rm km~s}^{-1}$ \citep{2013ARA&A..51..511K}. It is unlikely that this empirical relation between velocity dispersion and SMBH mass holds precisely at high redshift. However, this should be sufficiently accurate, since our results do not depend qualitatively on this parameter. 

We chose the outer radius, at which we set the $\dot{M}_{\rm out}$ boundary condition, to be 5 pc. This radius is consistent with observations of nearby AGN \citep{2013A&A...558A.149B}. Because the gas accretion rate is the only disk property that depends on the disk structure exterior to a given radius we have the freedom to excise, a posteriori, regions of the disk exterior to any given radius. Note that for some of our models, the disk can be more massive than the SMBH. However, in the most extreme cases the difference between disk mass and SMBH mass is only a factor of order unity, and our results do not depend strongly on the precise edge radius. Similarly, we neglect the mass of migrating compact objects in the disk, since this depends on disk properties both within and outside of a given radius. Additionally, if migration speeds are large, high accretion rates in compact objects can be realised with little compact object mass in the disk.

The TQM models originally used the dust and gas opacity tables from \citet{2003A&A...410..611S}. Since this opacity table only extends to $10^4$ Kelvin, we smoothly connect it to the OPAL opacity tables \citet{1996ApJ...464..943I} for our chosen metallicity (solar) in case temperatures exceed $10^4$ Kelvin.  We use these tabulated opacities in the star forming regions, but adopt an approximate combination of Kramers' free-free and bound-free opacities with electron scattering opacity in the inner stable region of the disk
\begin{equation}
\kappa=\kappa_{\rm es}+4\times10^{25}(1+X)(Z+0.001)\rho T^{-7/2},
\end{equation}
given in $\rm cm^2 ~ \rm g^{-1},$ and where $X=0.7381$ and $Z=0.0134$ are the hydrogen and metal mass fractions respectively \citep{2009ARA&A..47..481A} and $\kappa_{\rm es}=0.2(1+X)$ is the electron scattering opacity. Our reasons for this choice are illustrated by a comparison between our Figure \ref{fig: disk_par}, which displays disk parameters from a pair of our models, to the analogous Figure 6 in TQM. We find that using the \citet{2003A&A...410..611S} opacity tables for the inner regions of the disk leads to huge discontinuities in model parameters. These discontinuities can be multiple orders of magnitude, such as $\rho$ in TQM Figure 6, and are accompanied by sudden changes in gradients. We do not consider these changes physically meaningful; instead they are the result of applying realistic opacities to a simplified 1D disk model. It is possible that a disk model including vertical structure and radiative transport would ameliorate these issues. We find that by using a simplified opacity model, we limit discontinuities to the radius where we switch opacity prescriptions, and greatly reduce their impact compared to the discontinuities in TQM.

We note that \citet{2016ApJ...819L..17B} analysed TQM disk models using the \citet{2003A&A...410..611S} opacities and found migration traps associated with the discontinuities in the disk models. Our prescription reduces the surface density gradients and finds no migration traps. However, \citet{2016ApJ...819L..17B} also identified migration traps in the smooth \citet{2003MNRAS.341..501S} disk models, so it is plausible that migration traps may reemerge in disk models with fully consistent opacity prescriptions. 

We focus on the outer portion of the star forming region of the disk, outside of the opacity gap region.
We expect that there is insufficient time for star formation in the opacity gap region. To see this, note from Equation (\ref{eqn: frag_mass}) that when the opacity decreases with little change in temperature, the typical fragment mass drops by orders of magnitude to $\sim 1~ M_\odot$. The Kelvin-Helmholtz time for such stars is on the order of $\sim 30$ Myr. Considering that the opacity gap region occurs at $\sim 1 $pc, the Kelvin-Helmholtz time for protostars to begin fusion can be tens of thousands of dynamical timescales. Thus, it is likely that such protostars could be disrupted or accreted by more massive objects, which migrate through the disk much more quickly. Additionally, the doubling timescale for these stars via Bondi accretion is again many dynamical timescales. However, the gas density is sufficiently high in the opacity gap region that the doubling timescale for stars via Bondi accretion is less than the migration timescale ($t_{\rm mig}$) for these objects. Thus, we expect that, on rare occasions, stars could grow quickly to $300-500~ M_\odot$, limited by their luminosity and rotation, subsequently becoming black holes and contributing to disk structure in the same way as the stars that became black holes before entering the opacity gap region. 

The other parameter of interest is the feedback parameter $\epsilon$. The TQM models assume local feedback, so it useful to check if the model is compatible with migrating objects. The photon diffusion time can be approximated as 
\begin{equation}
t_{\rm diff} \approx h\tau/c \approx 3.26\tau \left(H/\rm pc\right)~ \rm yr.
\end{equation}
For migration timescales on the order of $\sim \rm Myr$, we find that the photon diffusion timescale is much shorter than migration timescales throughout our disk models. For example, in Figure \ref{fig: disk_par} we see that the largest $\tau$ is $\sim 10^5$ and the largest scale height is $\sim 1 ~\rm pc$, which occur at very different radii. Thus, the TQM feedback model is applicable to disks supported by feedback from migrating objects, at least to first order.

We assume that irradiation in the disk is primarily from accretion onto black holes, and that the gas supply to the migrating black holes is determined by local disk parameters. 
We investigate both $\epsilon=0.1$ and $\epsilon=0.4$ for moderately spinning and extremally spinning black holes. We assume these values of $\epsilon$ for three reasons, two physical and one practical. The first physical reason is that accretion onto black holes is far more efficient than fusion in stars. When migration timescales through the disk are much longer than stellar evolution timescales, which holds for most stars under our previous assumptions, an object will spend a majority of its time in the disk as a black hole rather than as a star. The second physical reason is that this interpretation of $\epsilon$ provides a straightforward way to understand why the opacity gap controls the accretion rate onto the SMBH: As opacity decreases, radiation feedback from accretion onto black holes in the disk would have comparatively little influence on gas far from the black hole, providing black holes with ample gas supply. Such a scenario may be able to feed embedded black holes at super-Eddington rates \citep{2014ApJ...796..106J} However, a detailed investigation into this scenario would require radiation hydrodynamics simulations..

From a practical standpoint, assuming that mass lost from the gas disk goes directly into black holes simplifies mass accounting. This is because stellar mass that does not contribute directly to black holes can be thought of as simply returning to the ambient gas density after supernovae. Additionally, small values of $\epsilon$ lead to enormous star formation rates that consume most of the gas flowing inwards via the disk. However, it is very difficult in such a scenario to supply the SMBH with gas at significant fraction of the Eddington rate, especially for low values of $\alpha$. Note that for a constant $\dot{M}$, lower viscosity implies higher surface density, which implies more star formation and leaves almost no gas to fuel the AGN. This could be partially mitigated, if we use $\epsilon \sim 10^{-3}$, by moving away from the $\alpha$-viscosity parameterization. Indeed, one expects additional torques in a mixed gas-stellar disk \citep{2011MNRAS.415.1027H}. However, we consider both $\alpha$-viscosity and $\epsilon \sim 10^{-1}$ to be reasonable stand-ins for the relevant physics.

We also assume negligible change in gas accretion rate due to accretion onto black holes in the gravitationally stable region of the disk. For the time being, this is a necessary assumption, as time-dependent disk models are beyond the scope of this work. This assumption is reasonable as far as pressure support is concerned, because higher gas densities and temperatures provide greater pressure in the stable disk regions than radiation pressure did in the unstable regions. This assumption is also justified after the fact by comparing the e-folding time of accretion onto black holes in the stable region of the disk to the migration time for our disk models, although this may not be the case for disks around more massive black holes. We perform this comparison in Section \ref{sec:Results}.

In order to gauge how sensitive our results are to our choices of model parameters, we consider different values of $\epsilon$, $\alpha$, and $\dot{M}_{\rm out}$. We choose $\epsilon = 0.1$ with $\alpha=\{0.2,0.25,0.3\}$, and $\epsilon=0.4$ with $\alpha=\{0.02,0.05,0.1\}$. For each combination of $\epsilon$ and $\alpha$ we choose one or two values of $\dot{M}_{\rm out}$ to investigate how accretion rate, and thus disk thickness, affect migration timescales and other results while holding other parameters constant. The range of $\alpha$ values was chosen to be consistent with a range of thin disk models, observations, and simulations \citep{2007MNRAS.376.1740K}. When quoting accretion rates without explicit units, we have normalised by the Eddington accretion rate, $\dot{M}_{\rm edd}=4\pi GM_\bullet m_h/(\eta\sigma_T c) $,  with $\eta=0.1$, where $m_h$ is the mass of atomic hydrogen.

\subsection{Summary} \label{subsec:summary}
Our process begins with a modified version of the TQM disk models, where we have used opacities valid over a wider range of temperatures to reduce discontinuities in the model; supposed that feedback occurs primarily by accretion of gas onto black holes in the disk; and then checked using many previous results that our overall picture is self-consistent. Our approach is limited in a number of ways. For example, we explicitly require that the disks have $Q=1$. Additionally, our model is agnostic with respect to the physical processes regulating accretion onto black holes embedded in the disk, applying the ansatz that precisely enough gas accretes onto embedded black holes to maintain $Q=1$. Similarly, there are may uncertainties and approximations associated with the migration, feedback, and accretion prescriptions that we employ. Because of these uncertainties and others, we do not attempt to create a disk model with completely self-consistent physics. 

We chose to investigate these disk models in order to assess the possibility of SMBH growth in the early universe through mergers with compact objects formed within their accretion disks. Because the models were constructed with this in mind, they have Eddington ratios than are considered typical for modern AGN. Accordingly, our disk models have larger scale heights at a given radius than most models of lower redshift AGN. Thinner disks would lead to both enhanced gap opening and faster migration in the absence of gap opening. Thus, we expect that the results presented here are not applicable to more slowly-accreting AGN.

\section{Results} \label{sec:Results}

\begin{figure}
\includegraphics[width=\columnwidth]{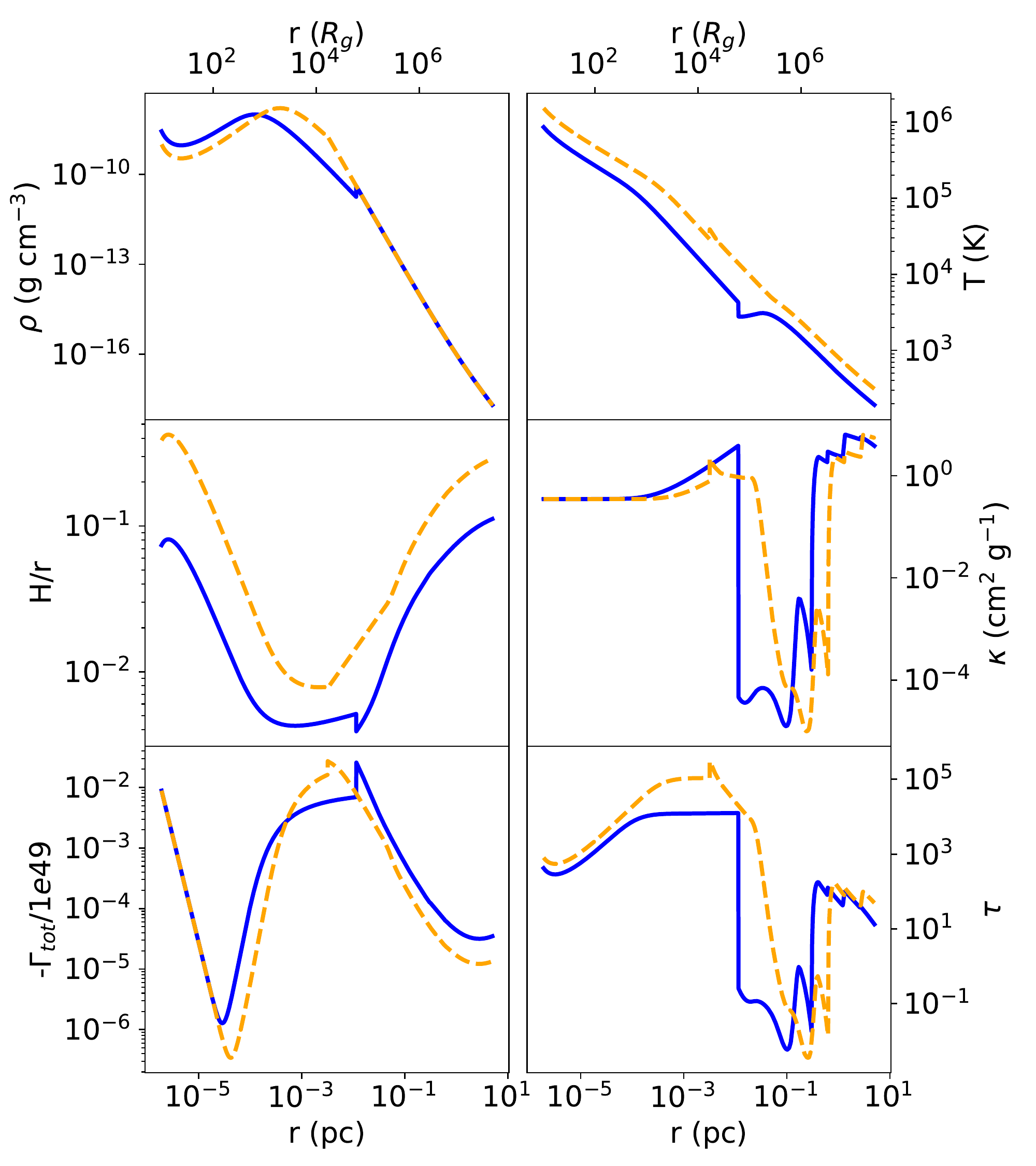}
\caption{Sample disk parameters for two of our models. Blue solid lines correspond to our model with $\epsilon=0.1$, $\alpha=0.2$ and $\dot{M}_{\rm out}=4.0$. Orange dashed lines correspond to our model with $\epsilon=0.4$, $\alpha=0.02$ and $\dot{M}_{\rm out}=7.0.$ From the top left: disk density; disk temperature; disk opening angle H/r; disk opacity; total torque from gravitational radiation and type-I migration torques;   disk optical depth.
Note that the torques we report are for a migrating object with mass $m=10~M_\odot$, but for $m\ll M_\bullet$, $\Gamma_{\rm tot}\propto m^2$, so this result can be extrapolated to other masses. }
\label{fig: disk_par}
\end{figure}

\begin{figure}
\includegraphics[width=\columnwidth,height=2.2in]{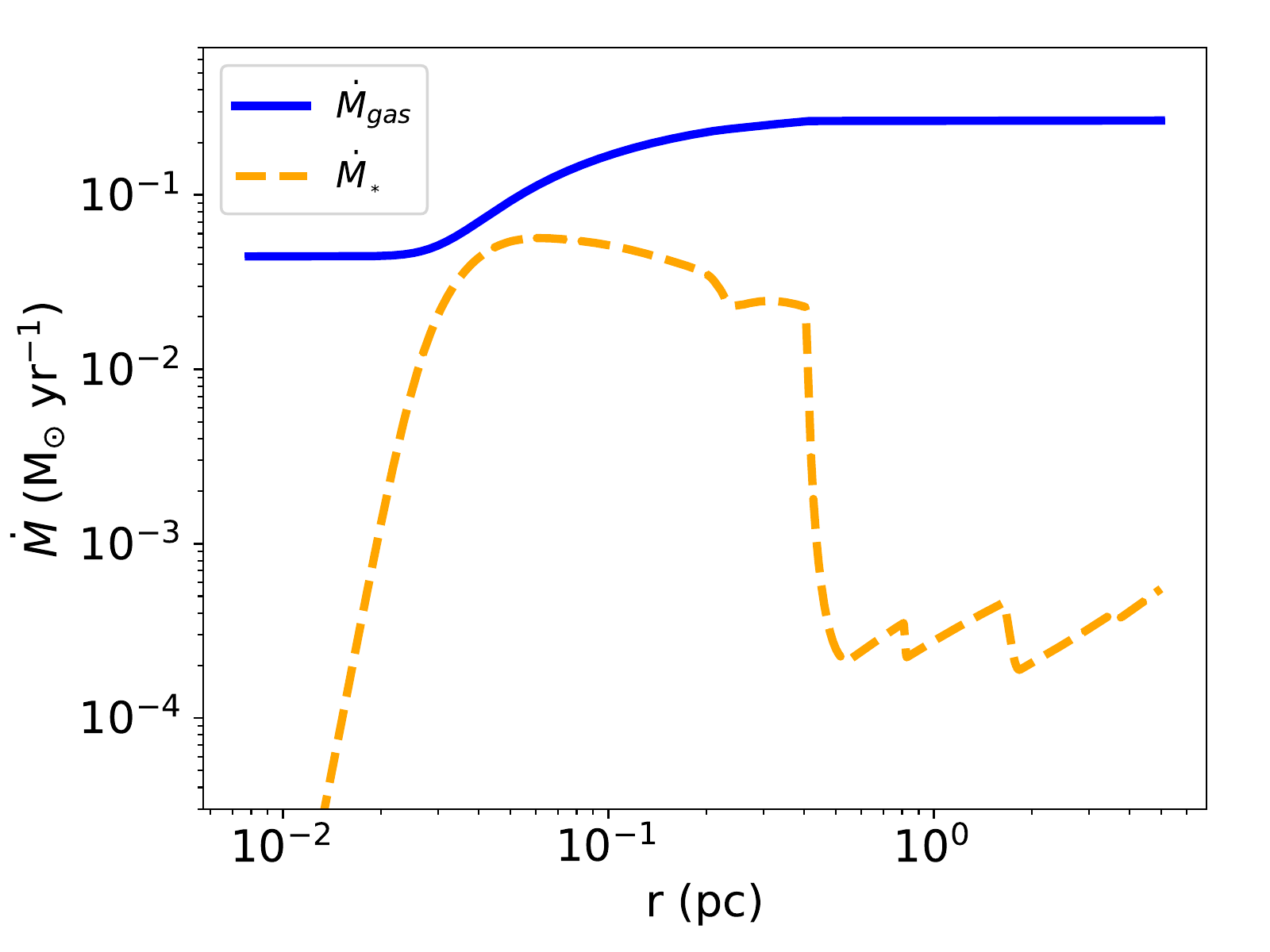}
\caption{Accretion rates in the star-forming region of our $\alpha=0.05$, $\dot{M}_{\rm out}=3.0$ disk model. The blue solid line plots the gas accretion rate through the disk, illustrating the change in gas accretion inwards towards the SMBH due to star formation and accretion onto stellar-mass black holes. The orange dashed line plots the growth rate of black holes in the disk $\dot{M}_*$.}
\label{fig: accretion}
\end{figure}

Figure \ref{fig: disk_par} depicts a number of disk parameters, both intrinsic and derived, over the full range of radii in our models. Note the `opacity gap' near temperatures of $10^3 \rm K$ corresponding to the sublimation of dust grains. 
This reduction in disk opacity means that embedded black holes can be fed at super-Eddington rates more easily, primarily limited by the higher opacity of gas as it accretes rather than the opacity of the disk. 
We present accretion rates in Figure \ref{fig: accretion}. Considering the bottom-left panel of Figure \ref{fig: disk_par}, it is evident that the migration of objects is always inward, regardless of the mass of the object, so migration traps will not occur in any of the disks we consider. Note the small dips in opacity in regions where $T \sim 100-1000 \rm K$, which occur as ice, volatile organics, or minerals such as troilite evaporate \citep{2003A&A...410..611S}. 

In Figure \ref{fig: accretion}, we present both the gas accretion rate $\dot{M}_{\rm gas}$ and the black hole growth rate $\dot{M}_*=\pi\dot{\Sigma}_*r^2$ for one of our models, where $\dot{\Sigma}_*$ is the accretion rate onto black holes in the disk. The other models are qualitatively similar, usually with slight differences in the location of the opacity gap or the magnitude of each curve. We chose this particular model because the change in $\dot{M}_{\rm gas}$ is easily visible, and
we calculate the accretion rate in black holes onto the central SMBH as the difference between $\dot{M}_{\rm out}$ and the value of $\dot{M}_{\rm gas}$ at the SMBH. It is clear that the star formation in the outer gap has little effect on the overall mass flowing into the AGN. However, these black holes are able to grow significantly in the mass gap. We note that the migration time is proportional to $m^{-1}$, so as mass is accreted onto black holes 
the inflowing mass constituted by black holes increases quadratically in the absence of gap opening.

\begin{figure}
\includegraphics[width=\columnwidth,height=2.2in]{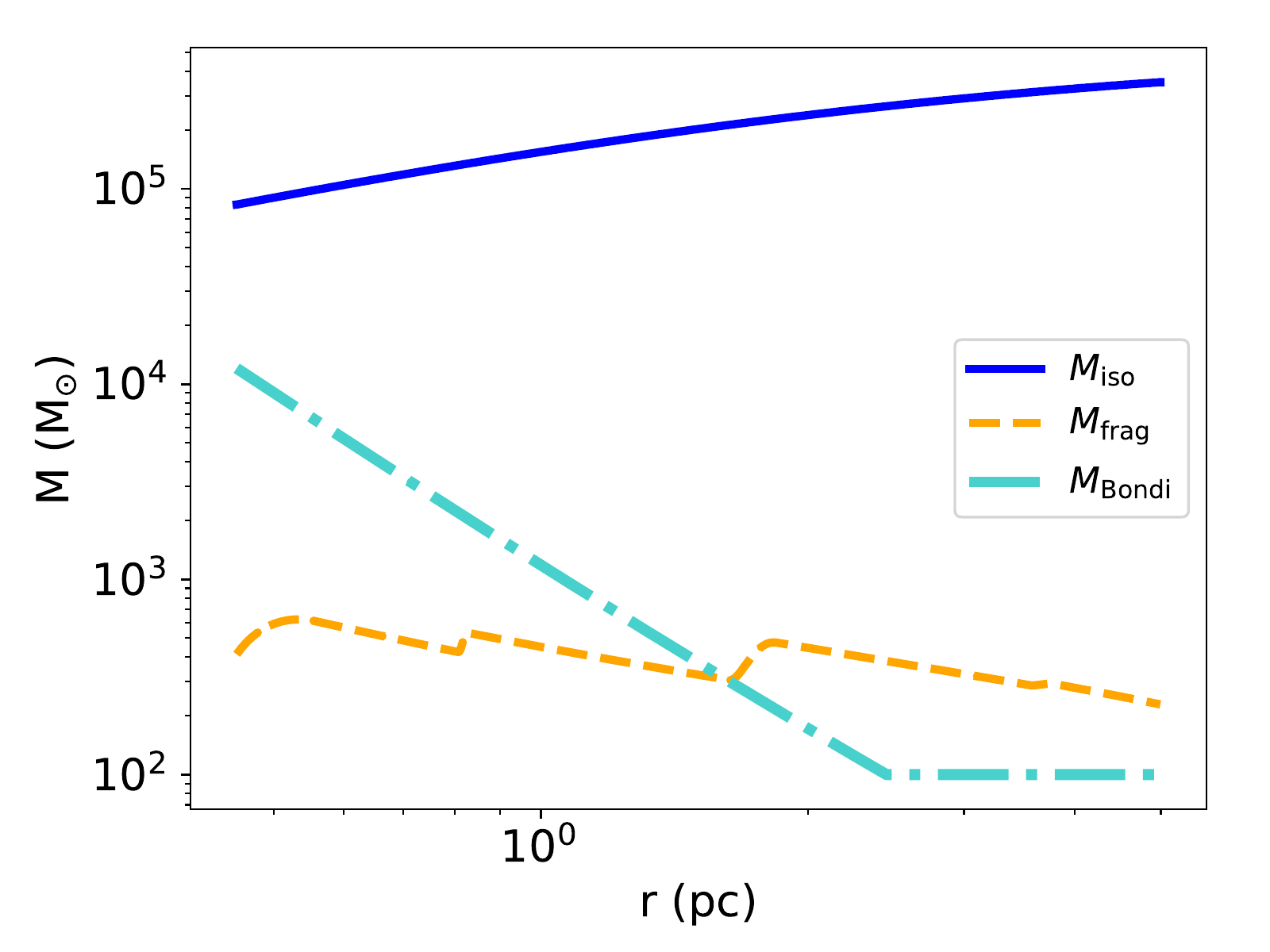}
\caption{Characteristic masses in the star-forming region of our $\alpha=0.05$, $\dot{M}_{\rm out}=3.0$ disk model. The blue solid line plots the isolation mass. The orange dashed line plots the fragment mass given by Equation (\ref{eqn: frag_mass}).
The dash-dotted line plots the equilibrium mass between Bondi accretion and mass loss, where we have set the minimum to 100 $M_\odot$ since our assumptions in calculating this balance are invalid below this mass. We expect other mass loss mechanisms to limit stellar masses to $\lesssim 500 ~M_\odot$. Other disk models are qualitatively similar to the one presented here. 
}
\label{fig: masses}
\end{figure}

We present sample characteristic masses in the star-forming region for the same disk model in Figure \ref{fig: masses}. These include the isolation mass, the initial fragment mass, and the equilibrium mass balancing Bondi-limited accretion and losses due to stellar winds. 
In the outer accretion disk, stellar masses are likely limited by the mass at which significant mass loss starts, likely between 50 and 100 $M_\odot$. We find that instead of being limited by the Eddington accretion rate, stars will likely be limited to masses below $\sim 300-500~ M_\odot$ by mass loss enhancements from rotation at high Eddington ratios. 

\begin{figure}
\includegraphics[width=\columnwidth,height=2.2in]{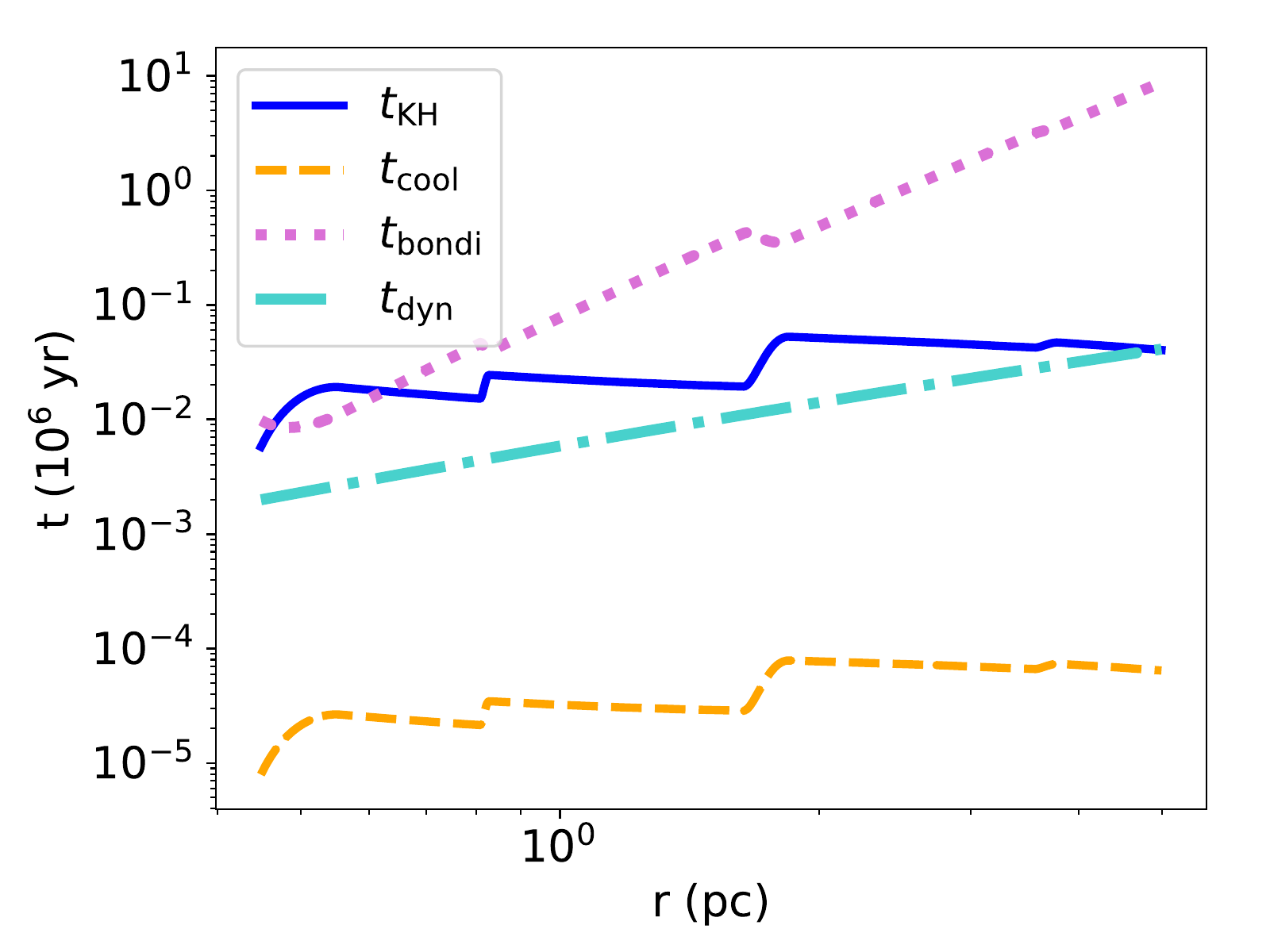}
\caption{Characteristic timescales in the star-forming region of our $\alpha=0.05$, $\dot{M}_{\rm out}=3.0$ disk model. The blue solid line indicates the Kelvin-Helmholtz timescale for fragments with mass given by Equation (\ref{eqn: frag_mass}). The orange dashed line indicates the cooling timescale in the disk. The dotted purple line indicates the Bondi doubling timescale of a fragment with mass given by Equation (\ref{eqn: frag_mass}). The turquoise dash-dotted line marks the dynamical timescale. The same hierarchy of timescales is shared between the disk models.
}
\label{fig: times}
\end{figure}

We must verify that various timescales relating to stellar accretion and evolution are self-consistent, as well as reasonable in the context of the stellar masses and accretion rates that we have predicted. Notably, all of the timescales we consider in Figure \ref{fig: times} are shorter than the main sequence lifetime of massive stars, ${\sim(2-3)}\times 10^6$ years \citep{1984ApJ...280..825B}, except for the doubling timescale at very large radii. Additionally, it is clear that protostars require a few orbits to reach the main sequence, but do not gain appreciable mass in this time except towards the inner edge of the star forming region. In most cases, fusion and therefore feedback processes can begin before accretion becomes significant. Additionally, the cooling timescale in the disk is much shorter than the dynamical timescale, as expected for a disk unstable to gravitational fragmentation. 

\begin{figure}
\includegraphics[width=\columnwidth,height=2.2in]{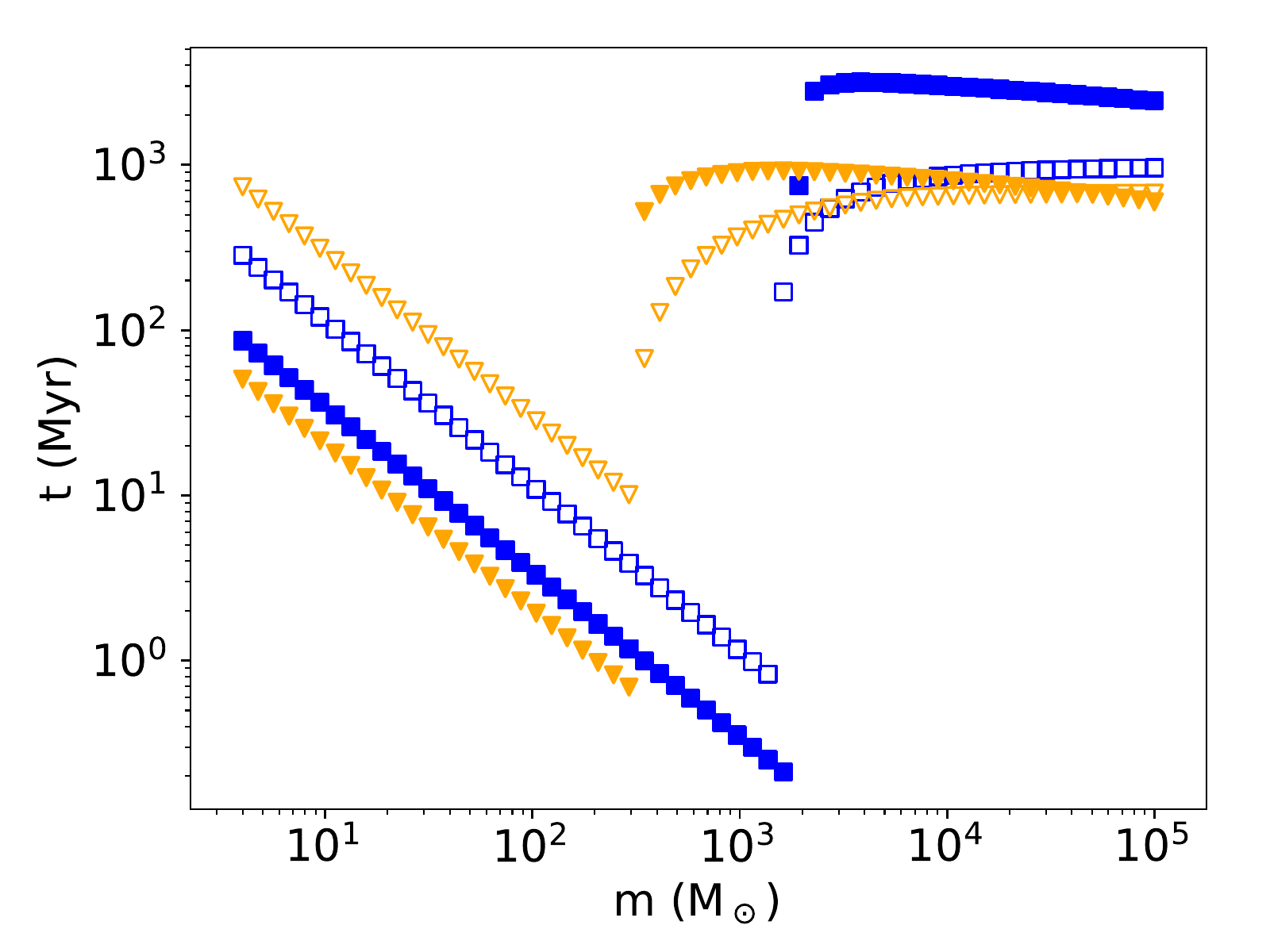}
\caption{Migration times for different migrating masses and different disk models. Hollow symbols represent the time required to migrate from the outer edge of the marginally stable disk region to the inner stable region. Filled symbols represent the amount of time required to migrate from the edge of the stable inner region to the central SMBH. Orange triangles correspond to the model with $\alpha=0.02, \dot{M}_{\rm out}=7.0$ and blue squares correspond to the model with $\alpha=0.3, \dot{M}_{\rm out}=3.0$. The large jumps in migration time correspond to masses that are able to open gaps in the disk.}
\label{fig: migration_time}
\end{figure}

In Figure \ref{fig: migration_time}, we present the migration times of objects through the disk for two disk models. We choose disk models with radically different viscosities, to illustrate the effects of gap opening, as well as to demonstrate some of the effects of disk thickness.
Recall from Equations (\ref{eqn: t_iso}), (\ref{eqn: t_adi}) and the expression for $\Gamma_0$, that thicker disks lead to slower migration. We separate migration time into two parts: The time to migrate through the inner stable region of the disk to the SMBH, and the time to migrate from the outer edge of the disk to the inner region. There are two trends for each disk model: as expected, migration time is inversely proportional to mass, up to the point where 
objects become massive enough to open gaps. The migration timescales in the outer disk are long enough for stars to become black holes and subsequently accrete. 
As we expect the lowest mass objects in the disk to be around $\sim 100 ~M_\odot$, almost all of the objects formed in the disk are able to migrate into the SMBH quickly enough to contribute to SMBH growth at high redshift.
Additionally, almost all stars born in these disks should have time to evolve into black holes before reaching the SMBH.

We also verify that our assumption of negligible accretion onto black holes in the inner region of the disk is appropriate. First, consider that the opacity in the stable region but more than $10^{-4}$ pc from the SMBH can be as many as 10 times the electron scattering opacity, so the e-folding time by accretion is on the order of $\sim 5\times10^8$ years for $\eta=0.1$. Then, considering a $\sim 100 ~M_\odot$ migrating object and consulting Figure \ref{fig: migration_time}, the total time spent by this object in the inner disk is $\sim 4\times 10^6$ yr, so changes in overall accretion rate in this region of the disk are negligible. 

\begin{table*}
\begin{centering}
\begin{tabular}{ccccccccccc}
$\epsilon$ & $\alpha$ & $\dot{M}_{\rm out}$ & $M_{ \rm frag, gap}$ & $M_{\rm frag}$ & $\tau_{100}$ & $\tau_{500}$ & $\dot{M}_{\rm gas}$ & $\dot{M}_{\rm bh}$ & $\Delta$ & $\Delta_{5}$ \\ \hline
0.4 & 0.02 & 7.0 & 169.72 & 435.85 & 31.51 & 949.88 & 1.29 & 5.71 & 0.15 & 0.47 \\
0.4 & 0.05 & 3.5 & 15.93 & 405.59 & 18.73 & 1117.32 & 0.79 & 2.71 & 0.23 & 0.59 \\
0.4 & 0.05 & 3.0 & 9.11 & 402.16 & 17.06 & 4517.14 & 0.50 & 2.50 & 0.17 & 0.50 \\
0.4 & 0.10 & 2.0 & 7.22 & 379.02 & 12.52 & 2.50 & 0.55 & 1.45 & 0.28 & 0.65 \\
0.1 & 0.20 & 4.0 & 0.64 & 379.26 & 11.81 & 2.36 & 0.24 & 3.76 & 0.06 & 0.24 \\
0.1 & 0.25 & 2.0 & 0.28 & 360.88 & 8.91 & 1.78 & 0.26 & 1.74 & 0.13 & 0.43 \\
0.1 & 0.30 & 1.5 & 0.58 & 351.90 & 7.94 & 1.59 & 0.31 & 1.19 & 0.21 & 0.57 \\
0.1 & 0.30 & 3.0 & 1.24 & 364.97 & 10.51 & 2.10 & 0.45 & 2.55 & 0.15 & 0.47 \\
\hline
\end{tabular}
\caption{Results for each model: $\epsilon$ is the efficiency parameter. $\alpha$ is the model viscosity parameter. $\dot{M}_{\rm out}$ is the gas accretion rate used as a boundary condition for the model at 5 pc.
$M_{\rm frag, gap}$ is the fragment mass, using Equation (\ref{eqn: frag_mass}), in the unstable region of the disk inside the opacity gap. $M_{\rm frag}$ is the median fragment mass, using Equation (\ref{eqn: frag_mass}), in the unstable region of the disk outside of the opacity gap.
$\tau_{\{100,500\}} $ are the total migration timescales in Myr for objects with masses 100 and 500 $M_\odot$. $\dot{M}_{\rm gas}$ is the gas accretion rate onto the central SMBH. $\dot{M}_{\rm bh}$ is the accretion rate, in black holes, onto the central SMBH. $\Delta$ is the fraction of the total Eddington-limited accretion rate comprised by gas. $\Delta_{5}$ is the same quantity if the accretion rate of black holes onto the SMBH has been overestimated by a factor of 5. Tabulated masses are in $M_\odot$, timescales in Myr, and accretion rates as a ratio to that of Eddington-limited accretion with $\eta=0.1$.}
\end{centering}
\label{tab:resuts}
\end{table*}

It is not practical to present plots of each parameter of interest for every disk model. However, we collect parameters of interest in Table \ref{tab:resuts}. $M_{\rm frag, gap}$ and $M_{\rm frag}$ are the median fragment masses given by Equation (\ref{eqn: frag_mass}) in the unstable region of the disk, inside and outside the opacity gap respectively. $M_{\rm frag, gap}$ provides a decent sense for fragment masses in the opacity gap region. Because the Kelvin-Helmholtz times for these 
lower-mass protostars are many times longer than those given by Equation (\ref{eqn:tkh}), collapse can take hundreds of dynamical times, making star formation in the opacity gap region unlikely.
On the other hand, the fragment masses $M_{\rm frag}$ are fairly consistent between disk models.  Although we do not expect star formation in the opacity gap region to alter disk structure significantly, stars that migrate into this region might grow to $300-500 ~M_\odot$, limited by rotation-enhanced winds. Such stars may collapse directly into black holes, unbinding little of their mass \citep{2001ApJ...550..372F}. In lower-$\alpha$ models, such black holes may open gaps at large distances from the SMBH, $\gtrsim 10^4 R_g$, precluding migration via gas or gravitational radiation torques. 

In order to understand the relevant migration timescales, we track the total migration times for objects with constant masses of $100 ~M_\odot, \tau_{100}$ and $500 ~M_\odot, \tau_{500}$. To calculate this quantity, we integrate $r/\dot{r}$ across our disk models. Since we have verified that the total torque never changes sign, this is an appropriate treatment. It is important to note how the migration timescale changes depending on object mass. For example, if the characteristic mass of migrating objects tends towards $\sim 500~ M_\odot$ or higher after accretion, these objects would open gaps, slowing down significantly.

Our method for calculating migration timescales when gap opening occurs likely results in an overestimate. Looking at the $H/r$ panel of Figure \ref{fig: disk_par} and recalling the dependence of the gap opening criterion on $H/r$, it is clear that once an object that can open a gap migrates inward, it can no longer open a gap, beginning to migrate quickly again. Gravitational interactions with other migrating objects may accelerate the rate at which gap-opening objects resume type-I migration. Consider a black hole massive enough to open a gap orbiting the SMBH, migrating inwards so slowly it can be considered to have a constant semi-major axis: Another black hole can then migrate towards it. In analogy to the hardening of binaries by 3-body encounters, the gap-opening object can move to a lower-energy orbit, transferring energy to the lighter object. Following this reasoning, a straightforward application of Equation (\ref{eqn: gap_migration}) may significantly overestimate the amount of time required for objects to migrate towards the SMBH is cases where gap opening occurs. Another possibility, when a type-I migrator encounters an object that has opened a gap, is that the type-I migrator is able to circumvent the gap on a horseshoe orbit similarly to the gap-crossing gas. In this case, we expect the migration timescale estimates to be unaffected. 

Similarly, let us consider the number and mass of black holes in the disk at one time, assuming temporarily that none have opened gaps. Since accretion rates and black hole masses are highest in the inner regions of the disk, consider the solid orange points in Figure \ref{fig: migration_time}. If $300 ~M_\odot$ is a characteristic mass for migrating black holes, each takes ${\sim 0.7 ~\rm Myr}$ to reach the SMBH, contributing about $1/1000$ of the $\sim 5.7~ \dot{M}_{\rm edd}$ accretion rate in black holes for the $\alpha=0.02$ model. In this case black holes in the disk total about 10\% the mass of the central SMBH. For comparison, consider the case $\alpha=0.1, \epsilon=0.4$ model if $500~ M_\odot$ black holes are characteristic representations for objects migrating through the disk. In this scenario, only $\sim 650$ black holes must occupy the disk, totalling about 8\% the mass of the SMBH. It is likely that our assumptions about gas torques damping eccentricities and inclinations, extrapolating the results of \citet{2008A&A...482..677C}, are invalid in this regime. However, this is a topic for future study. Note that more viscous disks can support migrating masses of more than $10^3~ M_\odot$, which could result in fewer than 100 black holes in the disk at a given time. This scenario also demonstrates that migrating black holes can make up a small fraction of the mass instantaneously orbiting the SMBH but dominate the accretion rate.

For now, we parameterise our uncertainty about the validity of our assumptions on migration in the inner disk by considering how our results change if we have overestimated the number of stellar mass black holes that reach the SMBH by some factor. Another possibility is that black holes in the disk can merge to alleviate the problem of large numbers of black holes, as each merger increases the accretion rate onto the SMBH, since $\tau \propto 1/m$. Mergers may be fairly common in the inner regions of the disk, where the number density of black holes is higher. Since we consider black holes which originate in the accretion disk, these black holes should have low relative speeds as they pass each other, which could facilitate merging. Excitation of orbit eccentricities also accelerates migration once gravitational radiation becomes a significant torque on migrating objects. 

We calculate $\Delta$, the change in accretion efficiency due to the presence of black holes in the accretion disk. Here,
\begin{equation}
\Delta = \frac{{\rm{min}}( \dot{M}_{\rm gas}, 1.0)}{ {\rm{min}}( \dot{M}_{\rm gas}, 1.0) + \dot{M}_{\rm bh}},
\end{equation}
since only this fraction of the total accretion rate results in the emission of photons. We define $\Delta_n$ by multiplying $\dot{M}_{\rm bh}$ by $1/n$ to evaluate how our estimate changes if we have overestimated the accretion rate of black holes onto the SMBH by a factor of $n$. This could occur, for example, if our assumptions of zero eccentricity and zero inclination orbits break down before torque due to gravitational waves becomes significant, or if black holes are ejected from the system.

We have verified many of the assumptions made during the construction of our disk models. For the purposes of growing high-z SMBHs, we have verified the necessary conditions that $t_{\rm kh} < t_{\rm mig}$ and that fragment masses in the disk are large enough to form stars as opposed to Jupiter-like objects. We find a range of potential factors that could decrease the radiative efficiency of SMBH growth. Because of the numerous unknowns in our treatment, such as the effects of feedback on migration torques and the dynamics of large number of migrating black holes in accretion disks, we consider it plausible that our values of $\Delta$ are overestimates. Even under these conservative considerations, the potential impact on high-z SMBH growth is significant.  

\section{Discussion} \label{sec:Discussion}

Let us turn to a more quantitative interpretation of our results. Consider our $\epsilon=0.4, \alpha=0.1$ case, which has both a sub-Eddington gas accretion rate and a comparatively mild accretion rate in black holes. In this case, a total accretion rate of about twice-Eddington is achieved even though the luminosity of the AGN would only be about half of the Eddington limit. This could effectively reduce the SMBH e-folding time by a factor of 2.

Very optimistically, one can interpret the results of other models as reducing the e-folding time by factors as low as 0.06. Such scenarios are extreme, and could not represent SMBH growth over long periods of time. However, such conditions might represent SMBH growth during a brief period of time, a few 10s of Myr, over which an SMBH is able to consume a large fraction of its disk emitting little radiation, possibly leaving behind rings of black holes and stars that formed but did not have time to migrate before the disk dissipated. If many black holes are ejected from the disk, if other stresses such as those from magnetic fields support the disk, or if global torques act to keep $Q\sim1$ through the marginally stable disk region, our $\Delta_{5}$ figures could represent a modification to accretion efficiency valid over long periods of time. Even in these situations, the accretion efficiency could change by factors $\sim 0.5$, which would significantly ease the growth of very massive AGN at high redshift. 

We expect that our analysis is also applicable to accretion disks around more massive SMBHs than the $4\times10^6 ~M_\odot$ case that we consider. The key consideration is how the Type-I migration timescale varies with central object mass, $\tau \propto M_\bullet$. For disks around less massive SMBHs, stars in the disk may not have time to exhaust their fuel before reaching the SMBH, being fully or partially disrupted, or undergoing stable Roche transfer \citep[e.g.][]{{2013MNRAS.434.2948D},{2017ApJ...844...75M}}, which would leave the accretion efficiency unchanged. However, growth via mergers with compact objects formed in accretion disks could potentially occur around more massive black holes without significant issue. Thus, although we do not think it possible to grow a seed black hole with $m \lesssim 80 ~M_\odot$ at $z \sim 20$ to $10^9~ M_\odot$ by $z\sim7$ using this mechanism alone, considering accretion of black holes formed in accretion disks can alleviate a number of tensions. 

Taking our $\epsilon=0.4,~ \alpha=0.1$ case as an example, an SMBH can grow from $10^6 ~M_\odot$ to $10^9~ M_\odot$ in only 155 Myr by accreting gas at $\sim 0.5$ times the Eddington rate. Considering only this change, the growth factor using $\eta=0.1$ for the first 345 Myr after $z \sim 20$, is approximately $2.1\times10^6$. If instead $\epsilon=0.1$ and $\alpha=0.2$ then
an SMBH can grow from $10^6~ M_\odot$ to $10^9 ~M_\odot$ in only 78 Myr by accreting gas at $\sim 0.25$ times the Eddington rate, facilitating its growth by a factor of $1.2\times10^7$ over the $\sim 500$ Myr between $z\sim20$ and $z\sim7$. Thus, it is possible that SMBHs with masses $\gtrsim 10^9 ~M_\odot$ grew from $\sim ~100~ M_\odot$ black holes via approximately steady accretion. Similarly, starting with a more massive seed black hole, mergers with black holes formed in accretion disks allow SMBHs to form at high redshift without invoking constant accretion. Note that if black holes are left orbiting the SMBH at the end of an accretion episode, they will experience drag as their orbits cross gas during the next accretion episode and and resume migration along with the next generation of black holes formed within the disk, analogous to the capture of stars into AGN disks discussed in \citet{1993ApJ...409..592A}.

\subsection{Observables} \label{subsec:obs}
Plentiful star formation in AGN disks has a number of other implications beyond growing SMBHs by $z\sim7$. These include, among others: metallicity enrichment of the AGN and surrounding galaxy \citep{1993ApJ...409..592A};  extreme mass ratio inspiral (EMRI) production; and transients such as superluminous supernovae. 

It is very difficult to measure the metallicities of the few AGN at $z\gtrsim6$ because most of the lines typically used to infer metallicity are redshifted to ranges difficult to observe from the ground. However, studies of quasar metallicity at $2.25 \lesssim z \lesssim 5.25$ have indicated that quasar broad line region (BLR) metallicities do not vary with redshift, although correlations between SMBH mass and quasar metallicity have been observed \citep{2018MNRAS.480..345X}. We note that if SMBH growth occurs significantly by accretion of compact objects formed in the disk, this AGN mass-metallicity relation follows naturally with little redshift dependence. Additionally, \citet{2018MNRAS.480..345X} found that BLR metallicities were systematically larger than host galaxy metallicities by factors larger than can be explained by uncertainties in their measurements. The $z \sim 7.5 $ quasar host galaxy J1342+0928 has a metallicity near solar \citep{2019ApJ...881...63N}, which would imply supersolar AGN metallicity. We suggest that AGN should naturally follow a mass-metallicity relationship regardless of redshift, provided the AGN disk can support star formation and those stars have time to reach the end of their lives ($M_\bullet \gtrsim 10^6 ~M_\odot$. The James Webb Space Telescope will facilitate direct metallicity measurement in $z \gtrsim 7$ AGN.

Based on the accretion rates of black holes discussed in the previous section, we can make illustrative estimates again using the $\epsilon=0.4$, $\alpha = 0.1$ model. Assuming that the black holes in the disk have masses $\sim 500~M_\odot$, we expect an EMRI rate of $\sim 10^{-4}~{\rm yr}^{-1}$ per AGN. Thus, space-based missions such as the Laser Interferometer Space Antenna \citep{2017arXiv170200786A} or the Chinese Taiji program \citep[e.g.,][]{2018arXiv180709495R} could detect such events with regularity. If we assume that the eccentricities and inclinations of stellar-origin black holes continue to damp as they migrate through the inner regions of the disk, many of these EMRI events would occur with nearly zero eccentricity. This would distinguish them from other EMRI mechanisms, which are expected to have large eccentricities when in the $\sim 10^{-4}-10^{-2}$~ Hz frequency range \citep{2007CQGra..24R.113A}. Capture of stars into the disk may also occur \citep[e.g.,][]{{2011MNRAS.413L..24M},{2011MNRAS.417L.103M}}, although we expect captured stars to have retrograde as well as prograde orbits, in contrast to the exclusively prograde orbits we expect from stars formed in-situ (assuming minimal perturbations). For retrograde orbits there exists the intriguing possibility of negative dynamical friction \citep{{2017ApJ...838..103P},{2019arXiv190601186G}} or other effects that could excite the eccentricities of the orbits.

As stars born in AGN accretion disks exhaust their fuel, we expect violent transients. As supernovae occur, we expect significant changes in AGN lightcurves. Typical Type-II supernovae rise in luminosity over the course of months, a signal that could be evident when superimposed over a stochastic AGN lightcurve. It is difficult to spectrally identify supernovae in AGN since Type-II supernovae generally have spectral lines that are similar to those in quasars. However, as the ejecta from these supernovae interact with the surrounding accretion disk, the flares could be significantly brighter and  longer lived, along the lines of the transients observed at lower redshift in \citet{2017MNRAS.470.4112G}. The rate of such flares in the lightcurves of high-z AGN could provide insight into star formation processes in the disk. AGN disks also provide a natural birthplace for pair-instability supernovae, as accretion onto stars could make up for the significant mass loss from stars above $\sim50 ~M_\odot$. Such events will likely be observable at $z \gtrsim 7$ by the James Webb Space Telescope \citep{2015ApJ...805...44S}. Additionally, pair-instability supernovae could account for why the flares noted in \citet{2017MNRAS.470.4112G} are more luminous than normal superluminous supernovae. 

For the purposes of growing SMBHs at high redshift, it is pertinent to study near-Eddington accretion. However, there are multiple reasons that it is unlikely that the method for growing SMBHs described here will be similarly effective in local AGN. Putting aside dependencies on SMBH mass, the timescale for type-I migration scales 
$\propto \left(H/r\right)^2$ \citep{2002ApJ...565.1257T}, as does the timescale for gas inflow due to viscous torques. Considering the simplified radiation-pressure supported disk models of \citet{1973A&A....24..337S},
$H \propto \dot{M}/\dot{M}_{\rm edd}$. Thus, for accretion disks with lower dimensionless accretion rates, stars may be more likely to reach the SMBH before reaching the end of their lives. Additionally, from eq. \ref{eq:g}, it is clear that gap opening is more prevalent in thinner disks. If gap opening becomes the norm rather than an exception for migrating objects, we expect significant departures from the results presented here. 

Finally, we can view our proposed mechanism in light of the So\l{}tan argument \citep{1982MNRAS.200..115S}, which connects the evolution of AGN luminosity and SMBH masses over cosmic time.  Recent estimates (e.g., \citealt{2004MNRAS.351..169M}) suggest that SMBH mass is accreted with an average efficiency of $\eta\approx 0.04-0.16$, and that a typical AGN accretes at a fraction of $\sim 0.1-1.7$ times the nominal Eddington rate.  Typical radiative efficiencies for pure gas accretion in this range of Eddington ratios are $\sim 0.06$ for slowly rotating black holes to $\sim 0.3$ for rapidly rotating holes.  Our finding that accretion of compact objects is potentially significant suggests that the derived average efficiencies imply a {\it gas} accretion efficiency that is a few times larger than $0.04-0.16$.  Thus we suggest that the spin parameters of SMBH could be systematically larger than normally inferred.  Note that observational constraints on $\eta$ inform us that the smaller values we find for $\Delta$ cannot represent a significant fraction of SMBH growth in the universe, although such disks may still exist infrequently. We note that lower radiative efficiencies can also be expected during super-Eddington accretion for other reasons  \citep[see, e.g.,][]{2019ApJ...880...67J}.

\section{Conclusions}\label{sec: conclusion}

We find that mergers between SMBHs and objects formed in or captured into the accretion disk can facilitate SMBH growth significantly beyond the standard Eddington-limited rate, even if the luminosity is capped at Eddington. This is therefore a potential channel for the production of $M\gtrsim 10^9~M_\odot$ black holes by $z\sim7$ from $\sim100 ~M_\odot$ seeds.

As previously suggested, stars can help explain the observed early enhancement of metallicity in disks \citep{1993ApJ...409..592A}, and we also suggest that in-situ star formation can explain the observed luminous long-timescale transients in AGN \citep{2017MNRAS.470.4112G}. This is also an interesting channel for EMRIs, which would have high enough secondary masses to produce events that can be detected easily and characterized precisely.

There are numerous unknowns in the physics relevant to our treatment. Examples include the balance between accretion and mass loss from stars, the torques in a mixed gas-stellar disks, and the thermal balance in disks that have as important elements both stars and accreting compact objects. Nonetheless, compact objects in disks around AGNs are likely to play important roles in many aspects of disk structure, supermassive black hole growth, and the production of gravitational wave sources. 

\section*{Acknowledgements}
The authors thank Nick Stone for many stimulating discussions on accretion disk models. The authors are also thankful to Brian Metzger, Yuri Levin, and Alberto Sesana for their comments on feedback, EMRIs, migration, and many other remarks that enhanced the clarity of this paper. The authors are grateful to Saavik Ford and Barry McKernan for useful discussions on the migration and orbital dynamics of black holes embedded in AGN disks. MCM is grateful for the hospitality of Perimeter Institute, where part of this work was carried out. Research at Perimeter Institute is supported in part by the Government of Canada through the Department of Innovation, Science and Economic Development Canada and by the Province of Ontario through the Ministry of Economic Development, Job Creation and Trade.



\bibliographystyle{mnras}
\bibliography{references} 

\begin{thebibliography}{}
\makeatletter
\relax
\def\mn@urlcharsother{\let\do\@makeother \do\$\do\&\do\#\do\^\do\_\do\%\do\~}
\def\mn@doi{\begingroup\mn@urlcharsother \@ifnextchar [ {\mn@doi@}
  {\mn@doi@[]}}
\def\mn@doi@[#1]#2{\def\@tempa{#1}\ifx\@tempa\@empty \href
  {http://dx.doi.org/#2} {doi:#2}\else \href {http://dx.doi.org/#2} {#1}\fi
  \endgroup}
\def\mn@eprint#1#2{\mn@eprint@#1:#2::\@nil}
\def\mn@eprint@arXiv#1{\href {http://arxiv.org/abs/#1} {{\tt arXiv:#1}}}
\def\mn@eprint@dblp#1{\href {http://dblp.uni-trier.de/rec/bibtex/#1.xml}
  {dblp:#1}}
\def\mn@eprint@#1:#2:#3:#4\@nil{\def\@tempa {#1}\def\@tempb {#2}\def\@tempc
  {#3}\ifx \@tempc \@empty \let \@tempc \@tempb \let \@tempb \@tempa \fi \ifx
  \@tempb \@empty \def\@tempb {arXiv}\fi \@ifundefined
  {mn@eprint@\@tempb}{\@tempb:\@tempc}{\expandafter \expandafter \csname
  mn@eprint@\@tempb\endcsname \expandafter{\@tempc}}}

\bibitem[\protect\citeauthoryear{{Amaro-Seoane}, {Gair}, {Freitag}, {Miller},
  {Mandel}, {Cutler}  \& {Babak}}{{Amaro-Seoane}
  et~al.}{2007}]{2007CQGra..24R.113A}
{Amaro-Seoane} P.,  {Gair} J.~R.,  {Freitag} M.,  {Miller} M.~C.,  {Mandel} I.,
   {Cutler} C.~J.,   {Babak} S.,  2007, \mn@doi [Classical and Quantum Gravity]
  {10.1088/0264-9381/24/17/R01}, \href
  {https://ui.adsabs.harvard.edu/abs/2007CQGra..24R.113A} {24, R113}

\bibitem[\protect\citeauthoryear{{Amaro-Seoane} et~al.,}{{Amaro-Seoane}
  et~al.}{2017}]{2017arXiv170200786A}
{Amaro-Seoane} P.,  et~al., 2017, arXiv e-prints, \href
  {https://ui.adsabs.harvard.edu/abs/2017arXiv170200786A} {p. arXiv:1702.00786}

\bibitem[\protect\citeauthoryear{{Appenzeller}}{{Appenzeller}}{1970}]{1970A&A.....5..355A}
{Appenzeller} I.,  1970, \aap, \href
  {https://ui.adsabs.harvard.edu/abs/1970A&A.....5..355A} {5, 355}

\bibitem[\protect\citeauthoryear{{Artymowicz}, {Lin}  \&
  {Wampler}}{{Artymowicz} et~al.}{1993}]{1993ApJ...409..592A}
{Artymowicz} P.,  {Lin} D.~N.~C.,   {Wampler} E.~J.,  1993, \mn@doi [\apj]
  {10.1086/172690}, \href
  {https://ui.adsabs.harvard.edu/abs/1993ApJ...409..592A} {409, 592}

\bibitem[\protect\citeauthoryear{{Asplund}, {Grevesse}, {Sauval}  \&
  {Scott}}{{Asplund} et~al.}{2009}]{2009ARA&A..47..481A}
{Asplund} M.,  {Grevesse} N.,  {Sauval} A.~J.,   {Scott} P.,  2009, \mn@doi
  [\araa] {10.1146/annurev.astro.46.060407.145222}, \href
  {https://ui.adsabs.harvard.edu/abs/2009ARA&A..47..481A} {47, 481}

\bibitem[\protect\citeauthoryear{{Bartos}, {Kocsis}, {Haiman}  \&
  {M{\'a}rka}}{{Bartos} et~al.}{2017}]{2017ApJ...835..165B}
{Bartos} I.,  {Kocsis} B.,  {Haiman} Z.,   {M{\'a}rka} S.,  2017, \mn@doi
  [\apj] {10.3847/1538-4357/835/2/165}, \href
  {https://ui.adsabs.harvard.edu/abs/2017ApJ...835..165B} {835, 165}

\bibitem[\protect\citeauthoryear{{Baruteau}, {Cuadra}  \& {Lin}}{{Baruteau}
  et~al.}{2011}]{2011ApJ...726...28B}
{Baruteau} C.,  {Cuadra} J.,   {Lin} D.~N.~C.,  2011, \mn@doi [\apj]
  {10.1088/0004-637X/726/1/28}, \href
  {https://ui.adsabs.harvard.edu/abs/2011ApJ...726...28B} {726, 28}

\bibitem[\protect\citeauthoryear{{Begelman}, {Volonteri}  \& {Rees}}{{Begelman}
  et~al.}{2006}]{2006MNRAS.370..289B}
{Begelman} M.~C.,  {Volonteri} M.,   {Rees} M.~J.,  2006, \mn@doi [\mnras]
  {10.1111/j.1365-2966.2006.10467.x}, \href
  {https://ui.adsabs.harvard.edu/abs/2006MNRAS.370..289B} {370, 289}

\bibitem[\protect\citeauthoryear{{Bellovary}, {Mac Low}, {McKernan}  \&
  {Ford}}{{Bellovary} et~al.}{2016}]{2016ApJ...819L..17B}
{Bellovary} J.~M.,  {Mac Low} M.-M.,  {McKernan} B.,   {Ford} K.~E.~S.,  2016,
  \mn@doi [\apjl] {10.3847/2041-8205/819/2/L17}, \href
  {https://ui.adsabs.harvard.edu/abs/2016ApJ...819L..17B} {819, L17}

\bibitem[\protect\citeauthoryear{{Bodenheimer} \& {Pollack}}{{Bodenheimer} \&
  {Pollack}}{1986}]{1986Icar...67..391B}
{Bodenheimer} P.,  {Pollack} J.~B.,  1986, \mn@doi [\icarus]
  {10.1016/0019-1035(86)90122-3}, \href
  {https://ui.adsabs.harvard.edu/abs/1986Icar...67..391B} {67, 391}

\bibitem[\protect\citeauthoryear{{Bond}, {Arnett}  \& {Carr}}{{Bond}
  et~al.}{1984}]{1984ApJ...280..825B}
{Bond} J.~R.,  {Arnett} W.~D.,   {Carr} B.~J.,  1984, \mn@doi [The
  Astrophysical Journal] {10.1086/162057}, \href
  {https://ui.adsabs.harvard.edu/abs/1984ApJ...280..825B} {280, 825}

\bibitem[\protect\citeauthoryear{{Bromm}}{{Bromm}}{2013}]{2013RPPh...76k2901B}
{Bromm} V.,  2013, \mn@doi [Reports on Progress in Physics]
  {10.1088/0034-4885/76/11/112901}, \href
  {https://ui.adsabs.harvard.edu/abs/2013RPPh...76k2901B} {76, 112901}

\bibitem[\protect\citeauthoryear{{Burtscher} et~al.,}{{Burtscher}
  et~al.}{2013}]{2013A&A...558A.149B}
{Burtscher} L.,  et~al., 2013, \mn@doi [\aap] {10.1051/0004-6361/201321890},
  \href {https://ui.adsabs.harvard.edu/abs/2013A%26A...558A.149B} {558, A149}

\bibitem[\protect\citeauthoryear{{Cai}, {Durisen}, {Michael}, {Boley},
  {Mej{\'\i}a}, {Pickett}  \& {D'Alessio}}{{Cai}
  et~al.}{2006}]{2006ApJ...636L.149C}
{Cai} K.,  {Durisen} R.~H.,  {Michael} S.,  {Boley} A.~C.,  {Mej{\'\i}a} A.~C.,
   {Pickett} M.~K.,   {D'Alessio} P.,  2006, \mn@doi [\apjl] {10.1086/500083},
  \href {https://ui.adsabs.harvard.edu/abs/2006ApJ...636L.149C} {636, L149}

\bibitem[\protect\citeauthoryear{{Chatzopoulos} \& {Wheeler}}{{Chatzopoulos} \&
  {Wheeler}}{2012}]{2012ApJ...748...42C}
{Chatzopoulos} E.,  {Wheeler} J.~C.,  2012, \mn@doi [\apj]
  {10.1088/0004-637X/748/1/42}, \href
  {https://ui.adsabs.harvard.edu/abs/2012ApJ...748...42C} {748, 42}

\bibitem[\protect\citeauthoryear{{Cresswell} \& {Nelson}}{{Cresswell} \&
  {Nelson}}{2008}]{2008A&A...482..677C}
{Cresswell} P.,  {Nelson} R.~P.,  2008, \mn@doi [\aap]
  {10.1051/0004-6361:20079178}, \href
  {https://ui.adsabs.harvard.edu/abs/2008A&A...482..677C} {482, 677}

\bibitem[\protect\citeauthoryear{{Dai} \& {Blandford}}{{Dai} \&
  {Blandford}}{2013}]{2013MNRAS.434.2948D}
{Dai} L.,  {Blandford} R.,  2013, \mn@doi [\mnras] {10.1093/mnras/stt1209},
  \href {https://ui.adsabs.harvard.edu/abs/2013MNRAS.434.2948D} {434, 2948}

\bibitem[\protect\citeauthoryear{{Davies}, {Miller}  \& {Bellovary}}{{Davies}
  et~al.}{2011}]{2011ApJ...740L..42D}
{Davies} M.~B.,  {Miller} M.~C.,   {Bellovary} J.~M.,  2011, \mn@doi [\apjl]
  {10.1088/2041-8205/740/2/L42}, \href
  {https://ui.adsabs.harvard.edu/abs/2011ApJ...740L..42D} {740, L42}

\bibitem[\protect\citeauthoryear{{Derdzinski}, {D'Orazio}, {Duffell}, {Haiman}
  \& {MacFadyen}}{{Derdzinski} et~al.}{2019}]{2019MNRAS.486.2754D}
{Derdzinski} A.~M.,  {D'Orazio} D.,  {Duffell} P.,  {Haiman} Z.,   {MacFadyen}
  A.,  2019, \mn@doi [\mnras] {10.1093/mnras/stz1026}, \href
  {https://ui.adsabs.harvard.edu/abs/2019MNRAS.486.2754D} {486, 2754}

\bibitem[\protect\citeauthoryear{{Devecchi} \& {Volonteri}}{{Devecchi} \&
  {Volonteri}}{2009}]{2009ApJ...694..302D}
{Devecchi} B.,  {Volonteri} M.,  2009, \mn@doi [\apj]
  {10.1088/0004-637X/694/1/302}, \href
  {https://ui.adsabs.harvard.edu/abs/2009ApJ...694..302D} {694, 302}

\bibitem[\protect\citeauthoryear{{Duffell}, {Haiman}, {MacFadyen}, {D'Orazio}
  \& {Farris}}{{Duffell} et~al.}{2014}]{2014ApJ...792L..10D}
{Duffell} P.~C.,  {Haiman} Z.,  {MacFadyen} A.~I.,  {D'Orazio} D.~J.,
  {Farris} B.~D.,  2014, \mn@doi [\apjl] {10.1088/2041-8205/792/1/L10}, \href
  {https://ui.adsabs.harvard.edu/abs/2014ApJ...792L..10D} {792, L10}

\bibitem[\protect\citeauthoryear{{D{\"u}rmann} \& {Kley}}{{D{\"u}rmann} \&
  {Kley}}{2015}]{2015A&A...574A..52D}
{D{\"u}rmann} C.,  {Kley} W.,  2015, \mn@doi [\aap]
  {10.1051/0004-6361/201424837}, \href
  {https://ui.adsabs.harvard.edu/abs/2015A&A...574A..52D} {574, A52}

\bibitem[\protect\citeauthoryear{{Ford} \& {McKernan}}{{Ford} \&
  {McKernan}}{2019}]{2019MNRAS.490L..42F}
{Ford} K.~E.~S.,  {McKernan} B.,  2019, \mn@doi [\mnras]
  {10.1093/mnrasl/slz116}, \href
  {https://ui.adsabs.harvard.edu/abs/2019MNRAS.490L..42F} {490, L42}

\bibitem[\protect\citeauthoryear{{Fryer}, {Woosley}  \& {Heger}}{{Fryer}
  et~al.}{2001}]{2001ApJ...550..372F}
{Fryer} C.~L.,  {Woosley} S.~E.,   {Heger} A.,  2001, \mn@doi [\apj]
  {10.1086/319719}, \href
  {https://ui.adsabs.harvard.edu/abs/2001ApJ...550..372F} {550, 372}

\bibitem[\protect\citeauthoryear{{Gammie}}{{Gammie}}{2001}]{2001ApJ...553..174G}
{Gammie} C.~F.,  2001, \mn@doi [\apj] {10.1086/320631}, \href
  {https://ui.adsabs.harvard.edu/abs/2001ApJ...553..174G} {553, 174}

\bibitem[\protect\citeauthoryear{{Goodman} \& {Tan}}{{Goodman} \&
  {Tan}}{2004}]{2004ApJ...608..108G}
{Goodman} J.,  {Tan} J.~C.,  2004, \mn@doi [The Astrophysical Journal]
  {10.1086/386360}, \href
  {https://ui.adsabs.harvard.edu/abs/2004ApJ...608..108G} {608, 108}

\bibitem[\protect\citeauthoryear{{Graham}, {Djorgovski}, {Drake}, {Stern},
  {Mahabal}, {Glikman}, {Larson}  \& {Christensen}}{{Graham}
  et~al.}{2017}]{2017MNRAS.470.4112G}
{Graham} M.~J.,  {Djorgovski} S.~G.,  {Drake} A.~J.,  {Stern} D.,  {Mahabal}
  A.~A.,  {Glikman} E.,  {Larson} S.,   {Christensen} E.,  2017, \mn@doi
  [\mnras] {10.1093/mnras/stx1456}, \href
  {https://ui.adsabs.harvard.edu/abs/2017MNRAS.470.4112G} {470, 4112}

\bibitem[\protect\citeauthoryear{{Gruzinov}, {Levin}  \& {Matzner}}{{Gruzinov}
  et~al.}{2019}]{2019arXiv190601186G}
{Gruzinov} A.,  {Levin} Y.,   {Matzner} C.~D.,  2019, arXiv e-prints, \href
  {https://ui.adsabs.harvard.edu/abs/2019arXiv190601186G} {p. arXiv:1906.01186}

\bibitem[\protect\citeauthoryear{{G{\"u}ltekin} et~al.,}{{G{\"u}ltekin}
  et~al.}{2009}]{2009ApJ...698..198G}
{G{\"u}ltekin} K.,  et~al., 2009, \mn@doi [\apj] {10.1088/0004-637X/698/1/198},
  \href {https://ui.adsabs.harvard.edu/abs/2009ApJ...698..198G} {698, 198}

\bibitem[\protect\citeauthoryear{{Hopkins} \& {Christiansen}}{{Hopkins} \&
  {Christiansen}}{2013}]{2013ApJ...776...48H}
{Hopkins} P.~F.,  {Christiansen} J.~L.,  2013, \mn@doi [\apj]
  {10.1088/0004-637X/776/1/48}, \href
  {https://ui.adsabs.harvard.edu/abs/2013ApJ...776...48H} {776, 48}

\bibitem[\protect\citeauthoryear{{Hopkins} \& {Quataert}}{{Hopkins} \&
  {Quataert}}{2011}]{2011MNRAS.415.1027H}
{Hopkins} P.~F.,  {Quataert} E.,  2011, \mn@doi [\mnras]
  {10.1111/j.1365-2966.2011.18542.x}, \href
  {https://ui.adsabs.harvard.edu/abs/2011MNRAS.415.1027H} {415, 1027}

\bibitem[\protect\citeauthoryear{{Hubeny}}{{Hubeny}}{1990}]{1990ApJ...351..632H}
{Hubeny} I.,  1990, \mn@doi [\apj] {10.1086/168501}, \href
  {https://ui.adsabs.harvard.edu/abs/1990ApJ...351..632H} {351, 632}

\bibitem[\protect\citeauthoryear{{Ida} \& {Makino}}{{Ida} \&
  {Makino}}{1992}]{1992Icar...96..107I}
{Ida} S.,  {Makino} J.,  1992, \mn@doi [\icarus]
  {10.1016/0019-1035(92)90008-U}, \href
  {https://ui.adsabs.harvard.edu/abs/1992Icar...96..107I} {96, 107}

\bibitem[\protect\citeauthoryear{{Iglesias} \& {Rogers}}{{Iglesias} \&
  {Rogers}}{1996}]{1996ApJ...464..943I}
{Iglesias} C.~A.,  {Rogers} F.~J.,  1996, \mn@doi [\apj] {10.1086/177381},
  \href {https://ui.adsabs.harvard.edu/abs/1996ApJ...464..943I} {464, 943}

\bibitem[\protect\citeauthoryear{{Inayoshi} \& {Haiman}}{{Inayoshi} \&
  {Haiman}}{2016}]{2016ApJ...828..110I}
{Inayoshi} K.,  {Haiman} Z.,  2016, \mn@doi [\apj]
  {10.3847/0004-637X/828/2/110}, \href
  {https://ui.adsabs.harvard.edu/abs/2016ApJ...828..110I} {828, 110}

\bibitem[\protect\citeauthoryear{{Jiang} \& {Goodman}}{{Jiang} \&
  {Goodman}}{2011}]{2011ApJ...730...45J}
{Jiang} Y.-F.,  {Goodman} J.,  2011, \mn@doi [\apj]
  {10.1088/0004-637X/730/1/45}, \href
  {https://ui.adsabs.harvard.edu/abs/2011ApJ...730...45J} {730, 45}

\bibitem[\protect\citeauthoryear{{Jiang}, {Stone}  \& {Davis}}{{Jiang}
  et~al.}{2014}]{2014ApJ...796..106J}
{Jiang} Y.-F.,  {Stone} J.~M.,   {Davis} S.~W.,  2014, \mn@doi [\apj]
  {10.1088/0004-637X/796/2/106}, \href
  {https://ui.adsabs.harvard.edu/abs/2014ApJ...796..106J} {796, 106}

\bibitem[\protect\citeauthoryear{{Jiang}, {Stone}  \& {Davis}}{{Jiang}
  et~al.}{2019}]{2019ApJ...880...67J}
{Jiang} Y.-F.,  {Stone} J.~M.,   {Davis} S.~W.,  2019, \mn@doi [\apj]
  {10.3847/1538-4357/ab29ff}, \href
  {https://ui.adsabs.harvard.edu/abs/2019ApJ...880...67J} {880, 67}

\bibitem[\protect\citeauthoryear{{Johnson}, {Goodman}  \& {Menou}}{{Johnson}
  et~al.}{2006}]{2006ApJ...647.1413J}
{Johnson} E.~T.,  {Goodman} J.,   {Menou} K.,  2006, \mn@doi [\apj]
  {10.1086/505462}, \href
  {https://ui.adsabs.harvard.edu/abs/2006ApJ...647.1413J} {647, 1413}

\bibitem[\protect\citeauthoryear{{Kanagawa}, {Tanaka}  \&
  {Szuszkiewicz}}{{Kanagawa} et~al.}{2018}]{2018ApJ...861..140K}
{Kanagawa} K.~D.,  {Tanaka} H.,   {Szuszkiewicz} E.,  2018, \mn@doi [\apj]
  {10.3847/1538-4357/aac8d9}, \href
  {https://ui.adsabs.harvard.edu/abs/2018ApJ...861..140K} {861, 140}

\bibitem[\protect\citeauthoryear{{King}, {Pringle}  \& {Livio}}{{King}
  et~al.}{2007}]{2007MNRAS.376.1740K}
{King} A.~R.,  {Pringle} J.~E.,   {Livio} M.,  2007, \mn@doi [\mnras]
  {10.1111/j.1365-2966.2007.11556.x}, \href
  {https://ui.adsabs.harvard.edu/abs/2007MNRAS.376.1740K} {376, 1740}

\bibitem[\protect\citeauthoryear{{Kolykhalov} \& {Syunyaev}}{{Kolykhalov} \&
  {Syunyaev}}{1980}]{1980SvAL....6..357K}
{Kolykhalov} P.~I.,  {Syunyaev} R.~A.,  1980, Soviet Astronomy Letters, \href
  {https://ui.adsabs.harvard.edu/abs/1980SvAL....6..357K} {6, 357}

\bibitem[\protect\citeauthoryear{{Kormendy} \& {Ho}}{{Kormendy} \&
  {Ho}}{2013}]{2013ARA&A..51..511K}
{Kormendy} J.,  {Ho} L.~C.,  2013, \mn@doi [\araa]
  {10.1146/annurev-astro-082708-101811}, \href
  {https://ui.adsabs.harvard.edu/abs/2013ARA&A..51..511K} {51, 511}

\bibitem[\protect\citeauthoryear{{Lamers} \& {Leitherer}}{{Lamers} \&
  {Leitherer}}{1993}]{1993ApJ...412..771L}
{Lamers} H. J.~G.~L.~M.,  {Leitherer} C.,  1993, \mn@doi [\apj]
  {10.1086/172960}, \href
  {https://ui.adsabs.harvard.edu/abs/1993ApJ...412..771L} {412, 771}

\bibitem[\protect\citeauthoryear{{Levin}}{{Levin}}{2007}]{2007MNRAS.374..515L}
{Levin} Y.,  2007, \mn@doi [\mnras] {10.1111/j.1365-2966.2006.11155.x}, \href
  {https://ui.adsabs.harvard.edu/abs/2007MNRAS.374..515L} {374, 515}

\bibitem[\protect\citeauthoryear{{Levin} \& {Beloborodov}}{{Levin} \&
  {Beloborodov}}{2003}]{2003ApJ...590L..33L}
{Levin} Y.,  {Beloborodov} A.~M.,  2003, \mn@doi [\apjl] {10.1086/376675},
  \href {https://ui.adsabs.harvard.edu/abs/2003ApJ...590L..33L} {590, L33}

\bibitem[\protect\citeauthoryear{{Low} \& {Lynden-Bell}}{{Low} \&
  {Lynden-Bell}}{1976}]{1976MNRAS.176..367L}
{Low} C.,  {Lynden-Bell} D.,  1976, \mn@doi [\mnras] {10.1093/mnras/176.2.367},
  \href {https://ui.adsabs.harvard.edu/abs/1976MNRAS.176..367L} {176, 367}

\bibitem[\protect\citeauthoryear{{Lyra}, {Paardekooper}  \& {Mac Low}}{{Lyra}
  et~al.}{2010}]{2010ApJ...715L..68L}
{Lyra} W.,  {Paardekooper} S.-J.,   {Mac Low} M.-M.,  2010, \mn@doi [\apjl]
  {10.1088/2041-8205/715/2/L68}, \href
  {https://ui.adsabs.harvard.edu/abs/2010ApJ...715L..68L} {715, L68}

\bibitem[\protect\citeauthoryear{{Maeder} \& {Meynet}}{{Maeder} \&
  {Meynet}}{2000}]{2000A&A...361..159M}
{Maeder} A.,  {Meynet} G.,  2000, \aap, \href
  {https://ui.adsabs.harvard.edu/abs/2000A&A...361..159M} {361, 159}

\bibitem[\protect\citeauthoryear{{Marconi}, {Risaliti}, {Gilli}, {Hunt},
  {Maiolino}  \& {Salvati}}{{Marconi} et~al.}{2004}]{2004MNRAS.351..169M}
{Marconi} A.,  {Risaliti} G.,  {Gilli} R.,  {Hunt} L.~K.,  {Maiolino} R.,
  {Salvati} M.,  2004, \mn@doi [\mnras] {10.1111/j.1365-2966.2004.07765.x},
  \href {https://ui.adsabs.harvard.edu/abs/2004MNRAS.351..169M} {351, 169}

\bibitem[\protect\citeauthoryear{{Masset} \& {Papaloizou}}{{Masset} \&
  {Papaloizou}}{2003}]{2003ApJ...588..494M}
{Masset} F.~S.,  {Papaloizou} J.~C.~B.,  2003, \mn@doi [\apj] {10.1086/373892},
  \href {https://ui.adsabs.harvard.edu/abs/2003ApJ...588..494M} {588, 494}

\bibitem[\protect\citeauthoryear{{McKernan}, {Ford}, {Yaqoob}  \&
  {Winter}}{{McKernan} et~al.}{2011a}]{2011MNRAS.413L..24M}
{McKernan} B.,  {Ford} K.~E.~S.,  {Yaqoob} T.,   {Winter} L.~M.,  2011a,
  \mn@doi [\mnras] {10.1111/j.1745-3933.2011.01024.x}, \href
  {https://ui.adsabs.harvard.edu/abs/2011MNRAS.413L..24M} {413, L24}

\bibitem[\protect\citeauthoryear{{McKernan}, {Ford}, {Lyra}, {Perets}, {Winter}
   \& {Yaqoob}}{{McKernan} et~al.}{2011b}]{2011MNRAS.417L.103M}
{McKernan} B.,  {Ford} K.~E.~S.,  {Lyra} W.,  {Perets} H.~B.,  {Winter} L.~M.,
   {Yaqoob} T.,  2011b, \mn@doi [\mnras] {10.1111/j.1745-3933.2011.01132.x},
  \href {https://ui.adsabs.harvard.edu/abs/2011MNRAS.417L.103M} {417, L103}

\bibitem[\protect\citeauthoryear{{McKernan}, {Ford}, {Lyra}  \&
  {Perets}}{{McKernan} et~al.}{2012}]{2012MNRAS.425..460M}
{McKernan} B.,  {Ford} K.~E.~S.,  {Lyra} W.,   {Perets} H.~B.,  2012, \mn@doi
  [\mnras] {10.1111/j.1365-2966.2012.21486.x}, \href
  {https://ui.adsabs.harvard.edu/abs/2012MNRAS.425..460M} {425, 460}

\bibitem[\protect\citeauthoryear{{McKernan}, {Ford}, {Kocsis}, {Lyra}  \&
  {Winter}}{{McKernan} et~al.}{2014}]{2014MNRAS.441..900M}
{McKernan} B.,  {Ford} K.~E.~S.,  {Kocsis} B.,  {Lyra} W.,   {Winter} L.~M.,
  2014, \mn@doi [\mnras] {10.1093/mnras/stu553}, \href
  {https://ui.adsabs.harvard.edu/abs/2014MNRAS.441..900M} {441, 900}

\bibitem[\protect\citeauthoryear{{McKernan} et~al.,}{{McKernan}
  et~al.}{2018}]{2018ApJ...866...66M}
{McKernan} B.,  et~al., 2018, \mn@doi [\apj] {10.3847/1538-4357/aadae5}, \href
  {https://ui.adsabs.harvard.edu/abs/2018ApJ...866...66M} {866, 66}

\bibitem[\protect\citeauthoryear{{McKernan}, {Ford}, {O'Shaughnessy}  \&
  {Wysocki}}{{McKernan} et~al.}{2019}]{2019arXiv190704356M}
{McKernan} B.,  {Ford} K.~E.~S.,  {O'Shaughnessy} R.,   {Wysocki} D.,  2019,
  arXiv e-prints, \href {https://ui.adsabs.harvard.edu/abs/2019arXiv190704356M}
  {p. arXiv:1907.04356}

\bibitem[\protect\citeauthoryear{{Metzger} \& {Stone}}{{Metzger} \&
  {Stone}}{2017}]{2017ApJ...844...75M}
{Metzger} B.~D.,  {Stone} N.~C.,  2017, \mn@doi [\apj]
  {10.3847/1538-4357/aa7a16}, \href
  {https://ui.adsabs.harvard.edu/abs/2017ApJ...844...75M} {844, 75}

\bibitem[\protect\citeauthoryear{{Mortlock} et~al.,}{{Mortlock}
  et~al.}{2011}]{2011Natur.474..616M}
{Mortlock} D.~J.,  et~al., 2011, \mn@doi [\nat] {10.1038/nature10159}, \href
  {https://ui.adsabs.harvard.edu/abs/2011Natur.474..616M} {474, 616}

\bibitem[\protect\citeauthoryear{{Nayakshin}, {Cuadra}  \&
  {Springel}}{{Nayakshin} et~al.}{2007}]{2007MNRAS.379...21N}
{Nayakshin} S.,  {Cuadra} J.,   {Springel} V.,  2007, \mn@doi [\mnras]
  {10.1111/j.1365-2966.2007.11938.x}, \href
  {https://ui.adsabs.harvard.edu/abs/2007MNRAS.379...21N} {379, 21}

\bibitem[\protect\citeauthoryear{{Novak} et~al.,}{{Novak}
  et~al.}{2019}]{2019ApJ...881...63N}
{Novak} M.,  et~al., 2019, \mn@doi [\apj] {10.3847/1538-4357/ab2beb}, \href
  {https://ui.adsabs.harvard.edu/abs/2019ApJ...881...63N} {881, 63}

\bibitem[\protect\citeauthoryear{{Paardekooper} \& {Papaloizou}}{{Paardekooper}
  \& {Papaloizou}}{2008}]{2008A&A...485..877P}
{Paardekooper} S.~J.,  {Papaloizou} J.~C.~B.,  2008, \mn@doi [\aap]
  {10.1051/0004-6361:20078702}, \href
  {https://ui.adsabs.harvard.edu/abs/2008A&A...485..877P} {485, 877}

\bibitem[\protect\citeauthoryear{{Paardekooper}, {Baruteau}, {Crida}  \&
  {Kley}}{{Paardekooper} et~al.}{2010}]{2010MNRAS.401.1950P}
{Paardekooper} S.~J.,  {Baruteau} C.,  {Crida} A.,   {Kley} W.,  2010, \mn@doi
  [\mnras] {10.1111/j.1365-2966.2009.15782.x}, \href
  {https://ui.adsabs.harvard.edu/abs/2010MNRAS.401.1950P} {401, 1950}

\bibitem[\protect\citeauthoryear{{Papaloizou}}{{Papaloizou}}{1973}]{1973MNRAS.162..169P}
{Papaloizou} J.~C.~B.,  1973, \mn@doi [\mnras] {10.1093/mnras/162.2.169}, \href
  {https://ui.adsabs.harvard.edu/abs/1973MNRAS.162..169P} {162, 169}

\bibitem[\protect\citeauthoryear{{Papaloizou} \& {Larwood}}{{Papaloizou} \&
  {Larwood}}{2000}]{2000MNRAS.315..823P}
{Papaloizou} J.~C.~B.,  {Larwood} J.~D.,  2000, \mn@doi [\mnras]
  {10.1046/j.1365-8711.2000.03466.x}, \href
  {https://ui.adsabs.harvard.edu/abs/2000MNRAS.315..823P} {315, 823}

\bibitem[\protect\citeauthoryear{{Park} \& {Bogdanovi{\'c}}}{{Park} \&
  {Bogdanovi{\'c}}}{2017}]{2017ApJ...838..103P}
{Park} K.,  {Bogdanovi{\'c}} T.,  2017, \mn@doi [\apj]
  {10.3847/1538-4357/aa65ce}, \href
  {https://ui.adsabs.harvard.edu/abs/2017ApJ...838..103P} {838, 103}

\bibitem[\protect\citeauthoryear{{Pepli{\'n}ski}, {Artymowicz}  \&
  {Mellema}}{{Pepli{\'n}ski} et~al.}{2008}]{2008MNRAS.386..179P}
{Pepli{\'n}ski} A.,  {Artymowicz} P.,   {Mellema} G.,  2008, \mn@doi [\mnras]
  {10.1111/j.1365-2966.2008.13046.x}, \href
  {https://ui.adsabs.harvard.edu/abs/2008MNRAS.386..179P} {386, 179}

\bibitem[\protect\citeauthoryear{{Peters}}{{Peters}}{1964}]{1964PhRv..136.1224P}
{Peters} P.~C.,  1964, \mn@doi [Physical Review] {10.1103/PhysRev.136.B1224},
  \href {https://ui.adsabs.harvard.edu/abs/1964PhRv..136.1224P} {136, 1224}

\bibitem[\protect\citeauthoryear{{Rafikov} \& {Slepian}}{{Rafikov} \&
  {Slepian}}{2010}]{2010AJ....139..565R}
{Rafikov} R.~R.,  {Slepian} Z.~S.,  2010, \mn@doi [\aj]
  {10.1088/0004-6256/139/2/565}, \href
  {https://ui.adsabs.harvard.edu/abs/2010AJ....139..565R} {139, 565}

\bibitem[\protect\citeauthoryear{{Ruan}, {Guo}, {Cai}  \& {Zhang}}{{Ruan}
  et~al.}{2018}]{2018arXiv180709495R}
{Ruan} W.-H.,  {Guo} Z.-K.,  {Cai} R.-G.,   {Zhang} Y.-Z.,  2018, arXiv
  e-prints, \href {https://ui.adsabs.harvard.edu/abs/2018arXiv180709495R} {p.
  arXiv:1807.09495}

\bibitem[\protect\citeauthoryear{{Secunda}, {Bellovary}, {Mac Low}, {Ford},
  {McKernan}, {Leigh}, {Lyra}  \& {S{\'a}ndor}}{{Secunda}
  et~al.}{2019}]{2019ApJ...878...85S}
{Secunda} A.,  {Bellovary} J.,  {Mac Low} M.-M.,  {Ford} K.~E.~S.,  {McKernan}
  B.,  {Leigh} N. W.~C.,  {Lyra} W.,   {S{\'a}ndor} Z.,  2019, \mn@doi [\apj]
  {10.3847/1538-4357/ab20ca}, \href
  {https://ui.adsabs.harvard.edu/abs/2019ApJ...878...85S} {878, 85}

\bibitem[\protect\citeauthoryear{{Semenov}, {Henning}, {Helling}, {Ilgner}  \&
  {Sedlmayr}}{{Semenov} et~al.}{2003}]{2003A&A...410..611S}
{Semenov} D.,  {Henning} T.,  {Helling} C.,  {Ilgner} M.,   {Sedlmayr} E.,
  2003, \mn@doi [\aap] {10.1051/0004-6361:20031279}, \href
  {https://ui.adsabs.harvard.edu/abs/2003A&A...410..611S} {410, 611}

\bibitem[\protect\citeauthoryear{{Shakura} \& {Sunyaev}}{{Shakura} \&
  {Sunyaev}}{1973}]{1973A&A....24..337S}
{Shakura} N.~I.,  {Sunyaev} R.~A.,  1973, \aap, \href
  {https://ui.adsabs.harvard.edu/abs/1973A&A....24..337S} {500, 33}

\bibitem[\protect\citeauthoryear{{Sirko} \& {Goodman}}{{Sirko} \&
  {Goodman}}{2003}]{2003MNRAS.341..501S}
{Sirko} E.,  {Goodman} J.,  2003, \mn@doi [\mnras]
  {10.1046/j.1365-8711.2003.06431.x}, \href
  {https://ui.adsabs.harvard.edu/abs/2003MNRAS.341..501S} {341, 501}

\bibitem[\protect\citeauthoryear{{Smidt}, {Whalen}, {Chatzopoulos}, {Wiggins},
  {Chen}, {Kozyreva}  \& {Even}}{{Smidt} et~al.}{2015}]{2015ApJ...805...44S}
{Smidt} J.,  {Whalen} D.~J.,  {Chatzopoulos} E.,  {Wiggins} B.,  {Chen} K.-J.,
  {Kozyreva} A.,   {Even} W.,  2015, \mn@doi [\apj]
  {10.1088/0004-637X/805/1/44}, \href
  {https://ui.adsabs.harvard.edu/abs/2015ApJ...805...44S} {805, 44}

\bibitem[\protect\citeauthoryear{{So\l{}tan}}{{So\l{}tan}}{1982}]{1982MNRAS.200..115S}
{So\l{}tan} A.,  1982, \mn@doi [\mnras] {10.1093/mnras/200.1.115}, \href
  {https://ui.adsabs.harvard.edu/abs/1982MNRAS.200..115S} {200, 115}

\bibitem[\protect\citeauthoryear{{Stewart} \& {Ida}}{{Stewart} \&
  {Ida}}{2000}]{2000Icar..143...28S}
{Stewart} G.~R.,  {Ida} S.,  2000, \mn@doi [\icarus] {10.1006/icar.1999.6242},
  \href {https://ui.adsabs.harvard.edu/abs/2000Icar..143...28S} {143, 28}

\bibitem[\protect\citeauthoryear{{Stone}, {Metzger}  \& {Haiman}}{{Stone}
  et~al.}{2017}]{2017MNRAS.464..946S}
{Stone} N.~C.,  {Metzger} B.~D.,   {Haiman} Z.,  2017, \mn@doi [\mnras]
  {10.1093/mnras/stw2260}, \href
  {https://ui.adsabs.harvard.edu/abs/2017MNRAS.464..946S} {464, 946}

\bibitem[\protect\citeauthoryear{{Tanaka} \& {Ward}}{{Tanaka} \&
  {Ward}}{2004}]{2004ApJ...602..388T}
{Tanaka} H.,  {Ward} W.~R.,  2004, \mn@doi [\apj] {10.1086/380992}, \href
  {https://ui.adsabs.harvard.edu/abs/2004ApJ...602..388T} {602, 388}

\bibitem[\protect\citeauthoryear{{Tanaka}, {Takeuchi}  \& {Ward}}{{Tanaka}
  et~al.}{2002}]{2002ApJ...565.1257T}
{Tanaka} H.,  {Takeuchi} T.,   {Ward} W.~R.,  2002, \mn@doi [\apj]
  {10.1086/324713}, \href
  {https://ui.adsabs.harvard.edu/abs/2002ApJ...565.1257T} {565, 1257}

\bibitem[\protect\citeauthoryear{{Thompson}, {Quataert}  \&
  {Murray}}{{Thompson} et~al.}{2005}]{2005ApJ...630..167T}
{Thompson} T.~A.,  {Quataert} E.,   {Murray} N.,  2005, \mn@doi [\apj]
  {10.1086/431923}, \href
  {https://ui.adsabs.harvard.edu/abs/2005ApJ...630..167T} {630, 167}

\bibitem[\protect\citeauthoryear{{Toomre}}{{Toomre}}{1964}]{1964ApJ...139.1217T}
{Toomre} A.,  1964, \mn@doi [The Astrophysical Journal] {10.1086/147861}, \href
  {https://ui.adsabs.harvard.edu/abs/1964ApJ...139.1217T} {139, 1217}

\bibitem[\protect\citeauthoryear{{Toyouchi}, {Hosokawa}, {Sugimura}, {Nakatani}
   \& {Kuiper}}{{Toyouchi} et~al.}{2019}]{2019MNRAS.483.2031T}
{Toyouchi} D.,  {Hosokawa} T.,  {Sugimura} K.,  {Nakatani} R.,   {Kuiper} R.,
  2019, \mn@doi [\mnras] {10.1093/mnras/sty3012}, \href
  {https://ui.adsabs.harvard.edu/abs/2019MNRAS.483.2031T} {483, 2031}

\bibitem[\protect\citeauthoryear{{Walsh}, {Morbidelli}, {Raymond}, {O'Brien}
  \& {Mandell}}{{Walsh} et~al.}{2011}]{2011Natur.475..206W}
{Walsh} K.~J.,  {Morbidelli} A.,  {Raymond} S.~N.,  {O'Brien} D.~P.,
  {Mandell} A.~M.,  2011, \mn@doi [\nat] {10.1038/nature10201}, \href
  {https://ui.adsabs.harvard.edu/abs/2011Natur.475..206W} {475, 206}

\bibitem[\protect\citeauthoryear{{Xu}, {Bian}, {Shen}, {Zuo}, {Fan}  \&
  {Zhu}}{{Xu} et~al.}{2018}]{2018MNRAS.480..345X}
{Xu} F.,  {Bian} F.,  {Shen} Y.,  {Zuo} W.,  {Fan} X.,   {Zhu} Z.,  2018,
  \mn@doi [\mnras] {10.1093/mnras/sty1763}, \href
  {https://ui.adsabs.harvard.edu/abs/2018MNRAS.480..345X} {480, 345}

\makeatother
\end{thebibliography}






\bsp	
\label{lastpage}
\end{document}